\documentstyle[epsf,tighten,preprint,amsfonts,eqsecnum,aps,prd]{revtex}

\begin{document}
\draft

\hyphenation{
mani-fold
mani-folds
geo-metry
geo-met-ric
}



\def\BbbR{{\Bbb R}}
\def\BbbZ{{\Bbb Z}}

\def\RPthree{{{\Bbb RP}^3}}

\def\sign{\mathop{\rm sign}\nolimits}

\def\eddfink{Eddington-Finkelstein}

\def\casehalf{{\case{1}{2}}}
\def\sfR{{\sf R}}

\def\rhat{{\hat r}}
\def\phat{{\hat p}}

\def\zr{{\frak r}}
\def\zp{{\frak p}}

\def\extder{\bbox{\delta}}

\def\outwidehat{\!\! \lower .2 em \hbox{$\widehat{\phantom{m}}$}}

\def\restmass{m}



\def\scriplus{{\cal I}^+}
\def\scriminus{{\cal I}^-}
\def\inought{i^0}

\def\gtildeta{{\tilde\Gamma}_\eta}

\def\bh{{\bf h}}

\def\bm{{\bf m}}
\def\bmplus{{\bf m}_+}
\def\bmminus{{\bf m}_-}
\def\bmplusminus{{\bf m}_\pm}

\def\bp{{\bf p}}
\def\bpplus{{\bf p}_+}
\def\bpminus{{\bf p}_-}
\def\bpplusminus{{\bf p}_\pm}

\def\bmbar{{\overline{\bf m}}}
\def\bmtilde{{\widetilde{\bf m}}}

\def\bpbar{{\overline{\bf p}}}
\def\bptilde{{\widetilde{\bf p}}}

\def\bpibar{{\overline{\bbox{\pi}}}}
\def\bpitilde{{\widetilde{\bbox{\pi}}}}

\def\bPibar{{\overline{\bbox{\Pi}}}}

\def\bptildeminus{{\tilde {\bf p}}_-}

\def\Rcut{R_{\rm cut}}

\def\mbar{{\overline{m}}}
\def\mtilde{{\widetilde{m}}}

\def\pbar{{\overline{p}}}
\def\ptilde{{\widetilde{p}}}

\preprint{\vbox{\baselineskip=12pt
\rightline{PP98--05}
\rightline{UF--RAP--97--19}
\rightline{WISC--MILW--97--TH--13}
\rightline{gr-qc/9708012}}}
\title{Hamiltonian spacetime dynamics with a spherical null-dust shell}
\author{Jorma Louko\footnote{On leave of absence from
Department of Physics, University of Helsinki.
Present address:
Max-Planck-Institut f\"ur Gravitations\-physik,
Schlaatzweg~1,
D--14473 Potsdam,
Germany.
Electronic address:
louko@aei-potsdam.mpg.de 
}}
\address{
Department of Physics,
University of Maryland,
College Park,
Maryland 20742--4111,
USA}
\author{Bernard F. Whiting\footnote{Electronic address:
bernard@bunyip.phys.ufl.edu}}
\address{
Department of Physics, University of Florida,
Gainesville, Florida 32611, USA}
\author{John L. Friedman\footnote{Electronic address:
friedman@thales.phys.uwm.edu}}
\address{
Department of Physics,
University of
Wisconsin--Milwaukee,\\
P.O.\ Box 413,
Milwaukee, Wisconsin 53201, USA}
\date{Published in 
{\it Phys.\ Rev.\ \rm D \bf 57}, 2279--2298 (1998)}
\maketitle
\begin{abstract}%
We consider the Hamiltonian dynamics of spherically symmetric Einstein
gravity with a thin null-dust shell, under boundary conditions that fix the
evolution of the spatial hypersurfaces at the two asymptotically flat
infinities of a Kruskal-like manifold. The constraints are eliminated
via a Kucha\v{r}-type canonical transformation and Hamiltonian reduction.
The reduced phase space ${\tilde\Gamma}$ consists of two disconnected
copies of~$\BbbR^4$, each associated with one direction of the shell
motion. The right-moving and left-moving test shell limits can be attached
to the respective components of ${\tilde\Gamma}$ as smooth
boundaries with topology~$\BbbR^3$. Choosing the right-hand-side and
left-hand-side masses as configuration variables provides a global
canonical chart on each component of~${\tilde\Gamma}$, and renders the
Hamiltonian simple, but encodes the shell dynamics in the momenta in a
convoluted way.  Choosing the shell curvature radius and the ``interior"
mass  as configuration variables renders the shell dynamics transparent in
an arbitrarily specifiable stationary gauge ``exterior" to the shell, but
the resulting local canonical charts do not cover the three-dimensional
subset of ${\tilde\Gamma}$ that corresponds to a horizon-straddling shell. 
When the evolution at the infinities is freed by introducing
parametrization clocks, we find on the unreduced phase space a global
canonical chart that completely decouples the physical degrees of freedom
from the pure gauge degrees of freedom.  Replacing one infinity by a
flat interior leads to analogous results, but with the reduced phase space
$\BbbR^2 \cup \BbbR^2$. The utility of the results  for quantization is
discussed. 
\end{abstract}

\narrowtext

\section{Introduction}
\label{sec:intro}

Spherically symmetric geometries have a long and useful history as a
physically interesting and technically vastly simplified arena for
gravitational physics. In vacuum, Einstein's theory with spherical symmetry
has no local degrees of freedom, and the reduced phase
space in the Hamiltonian formulation is finite dimensional.
Including an idealized, infinitesimally thin matter shell brings in an
additional finite number of degrees of freedom. Including a
continuous matter distribution generically yields a
(1+1)-dimensional field theory, with the exception of fields whose gauge
symmetries exclude spherically symmetric local degrees of freedom. A
familiar example of a field with such a gauge symmetry is the
electromagnetic field.

In this paper we consider spherically symmetric Einstein gravity coupled
to an infinitesimally thin null-dust shell. {}From the spacetime point of
view, the solutions to this system are well known (see, for example, Refs.\
\cite{redmount-shell,draythoft1,draythoft-cmp,clarkedray}), and they can be
easily obtained from a junction condition formalism that is general enough
to encompass null shells (see Ref.\ \cite{barr-is} and the references
therein). Our purpose is to explore the Hamiltonian structure of this
system, treating both the geometry and the shell as dynamical.
Among the extensive previous work on Hamiltonian approaches to spherically
symmetric geometries (for a selection in a variety of contexts, see
Refs.\
\cite{bcmn,unruh,thiemann1,thiemann2,kuchar1,thiemann3,thiemann4,%
LW2,cava1,cava2,lau1,marolf-boundary,lm,lou-win,%
kuns1,kuns2,kuns3,bro-kiefer,euclidean-refs,%
frolov,berezin88,fischler,hajicek,HKK,%
kraus-wilczek1,kraus-wilczek2,nakamura,dolgov,vakkuri-kraus,%
bicak-haji,kol-eard,FriLoWi,ansoldi+,%
lund,romano1,romano2}),
we follow most closely the canonical transformation techniques of
Kucha\v{r} \cite{kuchar1}.
Our main results can be concisely described as generalizing
the spherically symmetric vacuum Hamiltonian analysis of Ref.\
\cite{kuchar1} to accommodate a null-dust shell.

Finding a suitable action principle requires care. The shell stress-energy
tensor is a delta-distribution with support on the shell history, which is a 
hypersurface of codimension one. Einstein's equations for the
system therefore admit a consistent distributional interpretation
\cite{GT}, and the content of these equations is  captured by the junction
condition formalism of Barrabes and Israel \cite{barr-is}. We recover these
equations from a variational principle.  We take the shell action to be
that of a spherically symmetric thin cloud of radially-moving massless
relativistic point particles, and we  vary the total action independently
with respect to the gravitational variables and the shell variables. We
shall see that this variational principle can be made distributionally
consistent and that the variational equations do reproduce the correct
dynamics. Achieving this requires, however, a judicious choice of the
regularity properties of the metric.

We begin, in section~\ref{sec:metric}, by setting up the Hamiltonian
formulation of the system in the Arnowitt-Deser-Misner (ADM) gravitational
variables. The spacetime is taken to have Kruskal-like topology, with two
asymptotically flat infinities, and the spatial hypersurfaces are taken to
be asymptotic to hypersurfaces of constant Killing time at each spacelike
infinity. The Killing time evolution of the hypersurfaces is prescribed
independently at each infinity. We specify the regularity properties of the
gravitational variables, and demonstrate that the variational principle is
consistent and leads to the correct equations of motion.

In section \ref{sec:transformation} we perform a canonical transformation to
a new chart in which the constraints become exceedingly simple. Two of our
new variables are Hawking's quasilocal mass $M(r)$ and the two-sphere
curvature radius~$\sfR(r)$, just as in the vacuum analysis of Ref.\
\cite{kuchar1}. However, to maintain a consistent distributional
interpretation of the variables in the new chart, we are led to relate the
momentum conjugate to $M(r)$ to the \eddfink\ time whose constant value
hypersurface coincides with the classical shell history, and not to the
Killing time as in Ref.\ \cite{kuchar1}. The momentum conjugate to
$\sfR(r)$ needs to be modified accordingly. Remarkably, the canonical
transformation can then be chosen to leave the shell canonical pair
invariant. The transformation is mildly singular for geometries in which
the shell straddles a horizon, but it can be extended to this special case
in a suitable limiting sense.

In section \ref{sec:reduction} we eliminate the constraints by Hamiltonian
reduction. The reduced phase space ${\tilde\Gamma}$ turns out to have
dimension four. As the vacuum theory under our boundary conditions has a
two-dimensional reduced phase space \cite{kuchar1}, and as a test shell in
a fixed spherically symmetric background has a two-dimensional phase space,
this is exactly what one would have anticipated. We first obtain
canonical coordinates $(\bmplus,\bmminus,\bpplus,\bpminus)$ in which the
configuration variables $\bmplusminus$ are the Schwarzschild masses on the
two sides of the shell. The momenta $\bpplusminus$ can be interpreted as
the \eddfink\ time differences between the shell and the infinities, 
after introducing an appropriate correspondence between our spatial
hypersurfaces and hypersurfaces that are asymptotically null. The
configuration variables $\bmplusminus$ are constants of motion, while the
shell motion is indirectly encoded in the dynamics of~$\bpplusminus$. These
coordinates become singular for horizon-straddling shells, but a global
chart covering also this special case can be obtained by introducing
suitable new momenta. We find that ${\tilde\Gamma}$ consists of two
disconnected copies of~$\BbbR^4$, each associated with one direction of the
shell motion. The right-moving and left-moving test shell limits, in which
the shell stress-energy tensor vanishes, can be attached to the respective
components of ${\tilde\Gamma}$ as smooth boundaries with topology~$\BbbR^3$.

In section \ref{sec:gauges} we introduce on ${\tilde\Gamma}$ a local
canonical chart in which the shell motion becomes more transparent. Assuming
that the shell does not a straddle a horizon, the shell history
divides the spacetime into the ``interior," which contains a Killing horizon
bifurcation two-sphere, and the ``exterior," which does not. We choose the
configuration variables in the new chart to be the curvature radius of the
shell two-sphere and the interior mass, in an arbitrarily specifiable
stationary exterior coordinate system. One can argue that this yields a
Hamiltonian description of interest for an observer who scrutinizes the
shell motion from the exterior asymptotic region, especially if the
observer's ignorance of the interior asymptotic region is incorporated by
setting the interior contribution to the Hamiltonian to zero. We give three
examples of stationary exterior coordinate systems in which the Hamiltonian
can be found in closed form. Also, choosing the spatially flat exterior
gauge
\cite{painleve,gullstrand,israel-flat}, and performing a partial reduction
by setting the interior mass equal to a prescribed constant, we reproduce
the spatially flat shell Hamiltonian previously derived in Refs.\
\cite{kraus-wilczek1,FriLoWi} by different methods.

In section \ref{sec:para-times} we free the evolution of the spatial
hypersurfaces at the spacelike infinities by introducing parametrization
clocks. We find on the unreduced phase space a canonical chart in which the
physical degrees of freedom and pure gauge degrees of freedom are completely
decoupled, in full analogy with the vacuum analysis of Ref.\
\cite{kuchar1}. The pure gauge chart can be chosen so that the
configuration variables are the curvature radius of the two-sphere and the
\eddfink\ time, with the latter one appropriately interpreted across the
horizons.

In section \ref{sec:flat-int} we replace the Kruskal spatial topology
$S^2\times\BbbR$ by the spatial topology~$\BbbR^3$. The spacetime has then
just one asymptotic region, and when the equations of motion hold, the
spacetime interior to the shell is flat. As in the Kruskal case, we take the
asymptotic region to be asymptotically flat, and we prescribe the evolution
of the spatial hypersurfaces at the spacelike infinity. We then carry out
the canonical transformation and Hamiltonian reduction. Expectedly, the
reduced phase space turns out to consist of two disconnected copies
of~$\BbbR^2$, with only the counterpart of the pair $(\bmplus,\bpplus)$ of
the Kruskal theory surviving. 

We conclude in section \ref{sec:discussion} with a summary and a brief
discussion, including remarks on the potential utility of the results in
view of quantization. Some of the technical detail of the ADM dynamical
analysis is postponed to the
appendices.

We work in Planck units, $\hbar = c = G = 1$. Lowercase Latin tensor indices
$a,b,\dots$ are abstract spacetime indices. Dirac's delta-function is
denoted by~$\delta$, while $\extder$ denotes a variation. The curvature
coordinates $(T,R)$ for the Schwarzschild metric are coordinates in which
the metric reads
\begin{equation} ds^2 =
- (1 - 2M/R) dT^2
+ {(1 - 2M/R)}^{-1} dR^2
+ R^2 d\Omega^2
\ \ ,
\end{equation}
where $d\Omega^2$ is the metric on the unit two-sphere and $M$ is the
Schwarzschild mass. $T$~and $R$ are called respectively the Killing time
and the curvature radius.

\section{Metric formulation}
\label{sec:metric}

In this section we set up the Hamiltonian formulation for spherically
symmetric Einstein gravity coupled to a null-dust shell. We pay special
attention to the regularity of the gravitational variables and the global
boundary conditions.

\subsection{Bulk action}
\label{subsec:bulk}

Our spacetime geometry is given by the general spherically symmetric
Arnowitt-Deser-Misner (ADM) metric
\begin{equation}
ds^2 = - N^2 dt^2 + \Lambda^2 {(dr + N^r dt)}^2 +R^2 d\Omega^2
\ \ ,
\label{4-metric}
\end{equation}
where $d\Omega^2$ is the metric on the unit two-sphere, and $N$,
$N^r$, $\Lambda$, and $R$ are functions of the coordinates $t$ and $r$ only.
Partial derivatives with respect to $t$ and $r$ are denoted respectively by
overdot and prime,
$\dot{\phantom{a}} = \partial/\partial t$ and
${\phantom{a}}' = \partial/\partial r$. We take the spacetime metric to
be nondegenerate, and $N$, $\Lambda$, and $R$ to be positive.

The matter consists of an infinitesimally thin shell of dust
with a fixed total rest mass~$\restmass$, which we initially take to be
positive. Denoting the shell
history
by $r=\zr(t)$, the Lagrangian
matter action is
\begin{equation}
S^s_L =
- \restmass \int dt \,
\sqrt{ {\hat N}^2 - {\hat\Lambda}^2 {\left( \dot\zr +
{\widehat{N^r}} \right)}^2}
\ \ ,
\end{equation}
where a hat is used to denote the value of a variable at the shell. The
shell can be envisaged as a thin spherically symmetric cloud of
radially-moving massive relativistic point particles.

The Lagrangian gravitational action is obtained by performing integration
over the angles in the Einstein-Hilbert action, ${(16\pi)}^{-1}
\int d^4x \sqrt{-g}\,R$. Discarding a boundary term, the result is
\cite{bcmn,unruh,kuchar1,fischler,kraus-wilczek1,lund}
\begin{eqnarray}
S^g_L
= \int dt \int dr \,
\bigg[
&&
-N^{-1}
\left\{
R \bigl[ {\dot \Lambda}
- (\Lambda N^r)' \bigr]
({\dot R} - R' N^r )
+ \case{1}{2} \Lambda
{({\dot R} - R' N^r )}^2
\right\}
\nonumber
\\
&&
+ N
\left(
\Lambda^{-2} R R' \Lambda'
- \Lambda^{-1} R R''
- \case{1}{2} \Lambda^{-1} {R'}^{2}
+ \case{1}{2} \Lambda
\right) \bigg]
\ \ .
\label{S-lag}
\end{eqnarray}
The Lagrangian action of the coupled system is
\begin{equation}
S_L = S^g_L + S^s_L + \hbox{boundary terms}.
\end{equation}
We shall consider the regularity properties of the variables, the boundary
conditions, and boundary terms after passing to the Hamiltonian formulation.

The momenta conjugate to the configuration variables $\zr$, $\Lambda$,
and $R$ are
\begin{mathletters}
\begin{eqnarray}
\zp &=&
{
\restmass {\hat\Lambda}^2 {\left( \dot\zr +
{\widehat{N^r}} \right)}
\over
\sqrt{ {\hat N}^2 - {\hat\Lambda}^2 {\left( \dot\zr +
{\widehat{N^r}} \right)}^2}
}
\ \ ,
\\
\noalign{\smallskip}
P_{\Lambda} &=& - \frac{R}{N} (\dot R-N^rR')
\ \ ,
\label{PLambda}
\\
\noalign{\smallskip}
P_R &=&  -\frac{\Lambda}{N} (\dot R-N^r R')
-\frac{R}{N}
\left[ \dot\Lambda -
(N^r\Lambda)' \right]
\ \ .
\label{PR}
\end{eqnarray}
\end{mathletters}
A~Legendre transformation gives the Hamiltonian bulk action
\cite{fischler,kraus-wilczek1}
\begin{equation}
S_\Sigma =
\int dt \left[
\zp {\dot{\zr }} + \int dr \left (P_\Lambda \dot\Lambda +
P_R \dot R - NH - N^r H_r \right) \right]
\ \ ,
\label{S-ham}
\end{equation}
where the super-Hamiltonian constraint $H$ and the radial supermomentum
constraint $H_r$ contain both gravitational and matter contributions. In
the limit $\restmass\to0$, these constraints take the form
\begin{mathletters}
\label{metric-constraints}
\begin{eqnarray}
&&H =
{\Lambda P^2_\Lambda \over 2 R^2}
- { P_\Lambda P_R \over R}
+ { R R'' \over \Lambda}
- { R R' \Lambda' \over \Lambda^2}
+ { R'^2 \over 2\Lambda}
- {\Lambda \over 2}
+ {\eta \zp \over {\Lambda}} \, \delta (r-{\zr } )
\ \ ,
\label{superham}
\\
\noalign{\smallskip}
&&H_r = P_R R' - P'_\Lambda\Lambda - \zp\delta (r-\zr )
\ \ ,
\label{supermom}
\end{eqnarray}
\end{mathletters}
where $\eta := \sign(\zp)$. {}From now on, we shall work exclusively in
this zero rest mass limit, with the bulk action (\ref{S-ham}) and the
constraints~(\ref{metric-constraints}). As will be verified below, the
shell then consists of null dust.

The Hamiltonian constraint (\ref{superham}) is not differentiable in $\zp$
at $\zp=0$. As we shall verify below, an initial data set with nonzero
$\zp$ cannot evolve into a set with $\zp=0$
\cite{redmount-shell,draythoft1,draythoft-cmp,clarkedray,barr-is}. We assume
from now on that $\zp$ is nonzero: this breaks the phase space into the two
disconnected sectors $\eta=\pm1$. The limits $\zp\to0_\pm$ within each
sector will be addressed in subsection~\ref{subsec:reduction-no-shell}.

\subsection{Local equations of motion}
\label{subsec:local-eoms}

In the presence of a smooth matter distribution, one can assume the
spacetime metric to be smooth~($C^\infty$). In the idealized case of an
infinitesimally thin shell, the metric can be chosen continuous but not
differentiable at the shell \cite{barr-is,GT,israel-shell,MTW}. The issue
for us is to find smoothness assumptions that give a consistent variational
principle. We wish to make both the action (\ref{S-ham}) and its local
variations well defined and such that the resulting variational equations
are equivalent to Einstein's equations with a null-dust shell.

We follow the massive dust shell treatment of Ref.\ \cite{FriLoWi}. 
In contrast to the case of a massive dust shell, we shall find that the
smoothness conditions introduced in Ref.\ \cite{FriLoWi} make our null-dust
variational principle fully consistent.

As in Ref.\ \cite{FriLoWi}, we assume that the gravitational variables are
smooth functions of~$r$, with the exception that $N'$,
$(N^r)'$, $\Lambda'$, $R'$, $P_\Lambda$, and $P_R$ may have finite
discontinuities at isolated values of~$r$, and that the coordinate loci
of the discontinuities may be smooth functions of~$t$.  All the terms under
the $r$-integral in the action (\ref{S-ham}) are well defined in the
distributional sense. The most singular contributions are the explicit
matter delta-contributions in the constraints, and the implicit
delta-functions in $R''$ and~$P'_\Lambda$. All these delta-functions are
multiplied by continuous functions of~$r$. The remaining terms are at
worst  discontinuous in~$r$. The action is therefore well defined.

Local independent variations of the action with respect to the
gravitational and matter variables give the
constraint equations
\begin{mathletters}
\label{metric-constr-eoms}
\begin{eqnarray}
H &=& 0
\ \ ,
\label{metric-constr-eoms-ham}
\\
H_r &=& 0
\ \ ,
\end{eqnarray}
\end{mathletters}
and the dynamical equations
\begin{mathletters}
\label{metric-dyn-eoms}
\begin{eqnarray}
{\dot \Lambda}
&=&
N \left( {\Lambda P_\Lambda \over R^2}
    - {P_R \over R} \right)
+ {\left( N^r \Lambda \right)}'
\ \ ,
\label{Lambda-eom}
\\
{\dot R}
&=&
- {N P_\Lambda \over R}
+ N^r R'
\ \ ,
\label{R-eom}
\\
{\dot P}_\Lambda
&=&
{N \over 2}
\left[
- {P_\Lambda^2 \over R^2}
- \left({R' \over \Lambda}\right)^2
+ 1
+ {2\eta \zp \over \Lambda^2} \, \delta(r-\zr)
\right]
- {N' R R' \over \Lambda^2}
+ N^r P_\Lambda'
\ \ ,
\label{PLambda-eom}
\\
{\dot P}_R
&=&
N
\left[
{\Lambda P_\Lambda^2 \over R^3}
- {P_\Lambda P_R \over R^2}
- \left({R' \over \Lambda}\right)'
\right]
- \left( {N' R \over \Lambda}\right)'
+ {\left( N^r P_R \right)}'
\ \ ,
\label{PR-eom}
\\
{\dot {\zr}}
&=&
{ \eta {\hat N} \over {\hat \Lambda}} - {\widehat{N^r}}
\ \ ,
\label{rhat-eom}
\\
{\dot {\zp}}
&=&
\zp {\left[\left(
N^r- {\eta N \over \Lambda}\right)'\right]}^{\outwidehat}
\ \ .
\label{rhatmom-eom}
\end{eqnarray}
\end{mathletters}
As will be discussed in detail in appendix~\ref{app:dist-eoms-gen}, our
smoothness conditions imply that equations (\ref{metric-constr-eoms}) and
all save the last one of equations (\ref{metric-dyn-eoms}) have an
unambiguous distributional interpretation. Equation~(\ref{rhatmom-eom}), on
the other hand, is ambiguous: the right-hand side is a combination of
spatial derivatives evaluated at the shell, but these derivatives may be
discontinuous. We need to examine the dynamical content of the well-defined
equations, and the possibilities of interpreting
equation~(\ref{rhatmom-eom}).

A first observation from equation (\ref{rhat-eom}) is that the shell
history is tangent to the null vector $\ell^a$ whose components are
\begin{mathletters}
\label{ell-def}
\begin{eqnarray}
\ell^t &=& 1
\ \ ,
\\
\ell^r &=& \eta {\hat N}{\hat \Lambda}^{-1} - {\widehat{N^r}}
\ \ .
\end{eqnarray}
\end{mathletters}
For $\eta=1$ ($\eta=-1$), $\ell^a$~is the future null vector that points
towards relatively larger (smaller) values of~$r$. {}From the definition of
the shell stress-energy tensor, $\extder_g S_{\text{shell}}  = \casehalf
\int \sqrt{-g}\, d^4x \, T^{ab} \, \extder(g_{ab})$,
we find
\begin{equation}
T^{ab} = {\eta \zp \over 4\pi N^2 \Lambda^2 R^2} \, \ell^a \ell^b
\, \delta(r-\zr)
\ \ .
\label{shell-Tab}
\end{equation}
The shell is therefore null, with positive surface energy but
vanishing surface pressure \cite{barr-is}.
This confirms that the shell consists of null dust.

\begin{figure}
\begin{center}
\vglue 3 cm
\leavevmode
\epsfysize=6cm
\epsfbox{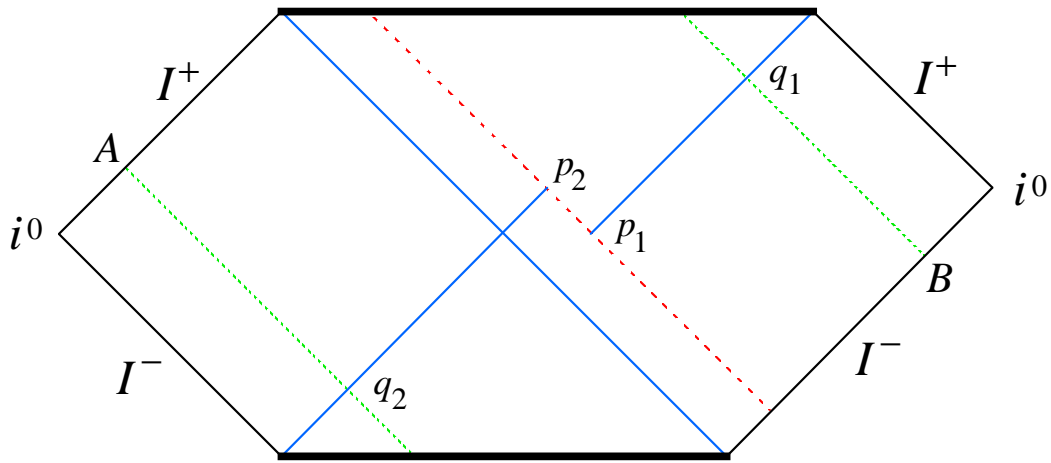}
\end{center}
\vskip 3 cm
\caption{The Penrose diagram for a spacetime in which the shell does not
straddle a horizon. The shell history is the dashed line passing
through points $p_1$ and~$p_2$. The shell has been taken left-moving, which
means $\eta=-1$, and to lie in the future of the left-going horizon, which
means that the right-hand-side Schwarzschild mass $\bmplus$ is greater than
the left-hand-side Schwarzschild mass~$\bmminus$. The diagrams for $\eta=1$
and/or $\bmplus<\bmminus$ are obtained through inversions of space or time
or both. The spacetime is uniquely determined by the values of~$\bmplus$,
$\bmminus$, and~$\eta$. A hypersurface of constant $t$ extends from the
left-hand-side $\inought$ to the right-hand-side~$\inought$, and the points
$A$ and $B$ indicate the ends of the asymptotically null hypersurface
introduced in section~\ref{sec:reduction}. The dotted lines are
hypersurfaces of constant null time ending respectively at $A$ and~$B$.
Point $B$ is here shown as being in the future of the shell history, but in
general it could be anywhere on the right-hand-side~$\scriminus$.}
\label{fig:nonhorizon}
\end{figure}

All solutions to the spherically symmetric Einstein equations with a
null-dust shell can be found from a sufficiently general junction
condition formalism
\cite{redmount-shell,draythoft1,draythoft-cmp,clarkedray,barr-is}.
On each side of the shell, the spacetime is locally part of the extended
Schwarzschild geometry. If the global structure of the spacetime is
Kruskal-like, with two  asymptotically flat infinities, there are only two
qualitatively different cases. First, if the shell is not static, the
junction is completely determined by continuity of the two-sphere radius at
the shell. The motion is clearly geodesic in each of the two geometries,
and the radius of the two-sphere serves as an affine parameter in either
geometry. The spacetime is either that shown in
figure~\ref{fig:nonhorizon}, or its time and/or space inverse. Second, if
the shell is static, the junction is along a common horizon, and the masses
must agree. The soldering is affine, meaning that the affine parameters
along the horizon with respect to the two geometries are affinely related;
however, as the stress-energy tensor of the shell is by assumption
nonvanishing, the bifurcation two-spheres on the two sides do not coincide.
The spacetime is either that shown in figure~\ref{fig:horizon} or its time
inverse.

Now, away from the shell, equations (\ref{metric-constr-eoms}) and
(\ref{metric-dyn-eoms}) are well known to be equivalent to Einstein's
equations. We shall investigate equations (\ref{metric-constr-eoms})
and (\ref{metric-dyn-eoms}) at the shell in detail in the appendices. 
The result is that, when combined with the fact that the geometry is locally
Schwarzschild on each side of the shell, the well-defined equations,
(\ref{metric-constr-eoms}) and (\ref{Lambda-eom})--(\ref{rhat-eom}), are
equivalent to the correct null-dust junction conditions at the shell. They
further imply that the right-hand side of (\ref{rhatmom-eom}) is
unambiguous, and that (\ref{rhatmom-eom}) is satisfied as an identity. Our
variational principle is therefore consistent, and it correctly reproduces
the motion of a null-dust shell.

A check on the consistency of our formalism is that the Poisson brackets of
our constraints can be shown to obey the radial hypersurface deformation
algebra \cite{teitel-dirac}, as in the absence of the shell, and as with a
massive dust shell \cite{FriLoWi}. We therefore have a Hamiltonian system
with first class constraints \cite{henn-teit-book}.

\subsection{Falloff}
\label{subsec:falloff}

What remains are the global boundary conditions. We take the coordinate $r$
to have the range $(-\infty,\infty)$, and  as $r\to\pm\infty$ we assume the
falloff \cite{kuchar1,beig-om}
\begin{mathletters}
\label{metricfall}
\begin{eqnarray}
\Lambda(t,r)
&=&
1 + M_\pm {|r|}^{-1} +
O^\infty({|r|}^{-1-\epsilon})
\ \ ,
\label{Lambdafall}
\\
R(t,r)
&=&
|r| + O^{\infty}({|r|}^{-\epsilon})
\ \ ,
\label{Rfall}
\\
P_\Lambda(t,r)
&=&
O^{\infty}({|r|}^{-\epsilon})
\ \ ,
\label{PLambdafall}
\\
P_R(t,r) & = &
O^{\infty}({|r|}^{-1-\epsilon})
\ \ ,
\label{PRfall}
\\
N(t,r)
&=&
N_\pm +
O^{\infty}({|r|}^{-\epsilon})
\ \ ,
\label{Nfall}
\\
N^{r}(t,r)
&=&
O^{\infty}({|r|}^{-\epsilon})
\ \ ,
\label{Nrfall}
\end{eqnarray}
\end{mathletters}
where $M_\pm$ and $N_\pm$ are functions of~$t$, and $\epsilon$ is a
parameter that can be chosen freely in the range $0<\epsilon\le1$. Here,
$O^{\infty}$ stands for a term that falls off as $r\to\pm\infty$ as its
argument, and whose derivatives with respect to $r$ and $t$ fall off
accordingly. These conditions imply that the asymptotic regions associated
with $r\to\pm\infty$ are asymptotically flat, with the constant $t$
hypersurfaces asymptotic to hypersurfaces of constant Minkowski time.
$N_\pm$ are the rates at which the asymptotic Minkowski times evolve with
respect to the coordinate time~$t$. When the equations of motion hold,
$M_\pm$ are time-independent and equal to the Schwarzschild masses.

In the variational principle, we take $N_\pm$ to be prescribed functions
of~$t$,  but leave $M_\pm$ free. The
appropriate total action then reads \cite{kuchar1}
\begin{mathletters}
\label{total-metric-action}
\begin{equation}
S = S_\Sigma + S_{\partial\Sigma}
\ \ ,
\end{equation}
where the boundary action is
\begin{equation}
S_{\partial\Sigma}
=
- \int dt \, ( N_+ M_+ + N_- M_- )
\ \ .
\end{equation}
\end{mathletters}
The global structure of the spacetime is Kruskal-like, with two
asymptotically flat asymptotic regions. The classical solutions under these
boundary conditions are precisely those described above and shown in
figures \ref{fig:nonhorizon} and~\ref{fig:horizon}.

\section{Canonical transformation}
\label{sec:transformation}

In this section we find a new canonical chart in which the constraints
become exceedingly simple. Away from the shell, our treatment closely
follows that given by Kucha\v{r} in the vacuum case \cite{kuchar1}. The new
elements arise mainly from patching the two vacuum regions together at the
shell.

Our canonical transformation turns out to be mildly singular when the
masses on the two sides of the shell agree and the shell straddles a common
horizon. We first perform the transformation, in
subsection~\ref{subsec:trans-nonhorizon}, assuming that this special case
has been excluded. We then argue, in subsection~\ref{subsec:trans-horizon},
that the transformation can be extended to the special case in a suitable
limiting sense.

\subsection{Shell not on a horizon}
\label{subsec:trans-nonhorizon}

In the vacuum theory, Kucha\v{r} \cite{kuchar1} found
a transformation from the canonical chart $\left(\Lambda, R,
P_\Lambda, P_R\right)$ to the new canonical chart
$\left(M, \sfR, P_M, P_\sfR\right)$ defined by
\begin{mathletters}
\label{trans}
\begin{eqnarray}
M &:=& \case{1}{2} R (1-F)
\ \ ,
\label{trans-M}
\\
\sfR &:=& R
\ \ ,
\label{trans-sfR}
\\
P_M &:=& R^{-1} F^{-1} \Lambda P_\Lambda
\ \ ,
\label{trans-PM}
\\
P_\sfR &:=&
P_R
- \case{1}{2} R^{-1} \Lambda P_\Lambda
- \case{1}{2} R^{-1} F^{-1} \Lambda P_\Lambda
\nonumber
\\
&& - R^{-1} \Lambda^{-2} F^{-1}
\left[
{(\Lambda P_\Lambda)}' (RR')
- (\Lambda P_\Lambda) {(RR')}'
\right]
\ \ ,
\label{trans-PsfR}
\end{eqnarray}
\end{mathletters}
where
\begin{equation}
F :=
{\left( {R' \over \Lambda} \right)}^2
- {\left( { P_\Lambda \over R} \right)}^2
\ \ .
\label{F-def}
\end{equation}
When the equations of motion hold, $M$ is independent of both $r$ and~$t$,
and its value is just the Schwarzschild mass. Similarly, when the equations
of motion hold, we have $P_M = -T'$, where $T$ is the Killing time.
The vacuum constraints can be written as a linear combination of
$M'$ and~$P_\sfR$, and the dynamical content of
the theory becomes transparent.

In the presence of our null shell, the variables $\left(M, \sfR,
P_M, P_\sfR\right)$ become singular at the shell. To see this,
consider a classical solution in which the shell
history does not lie on a horizon.
As $M$ is discontinuous at the shell, ${\dot{M}}$
contains at the shell a delta-function in~$r$. As $P_M$ is
discontinuous at the shell, the product $P_M {\dot{M}}$ is
ambiguous. One therefore does not expect the chart $\left(M, \sfR, P_M,
P_\sfR\right)$ to be viable in the presence of the shell.

To overcome this difficulty, we keep $M$ and $\sfR$ but replace the momenta
by ones that are smoother across the shell. We define first
\cite{kuchar1}
\begin{equation}
F_\pm := {R' \over \Lambda} \pm
{P_\Lambda \over R}
\end{equation}
and
\begin{equation}
F_{\pm \eta}:= {R' \over \Lambda} \pm \eta {P_\Lambda \over R}
\ \ .
\end{equation}
Note that $F= F_+ F_- = F_\eta F_{-\eta}$. When the equations of motion
hold,
$F_+$ vanishes on the leftgoing branch(es) of the horizon and
$F_-$ vanishes on the rightgoing branch(es) of the horizon. It follows
that $F_{-\eta}$ is nonvanishing on the horizon that the shell crosses.
Now let
\begin{mathletters}
\label{Pi-def}
\begin{eqnarray}
\Pi_M &:=& P_M + \eta F^{-1} R'
\nonumber
\\
&=&
\eta \Lambda F_{-\eta}^{-1}
\ \ ,
\label{PiM-def}
\\
\Pi_\sfR &:=& P_\sfR - \eta F^{-1} M'
\nonumber
\\
&=&
P_R + \eta R {\left( \ln \left| F_{-\eta} \right| \right)}'
+ {\eta\Lambda \over 2}\left( F_{-\eta} - F_{-\eta}^{-1} \right)
\label{PisfR-def}
\ \ .
\end{eqnarray}
\end{mathletters}
When the equations of motion hold, equation (\ref{PiM-def}) shows that
\begin{equation}
\Pi_M = -(T - \eta r^*)'
\ \ ,
\label{PiMprime}
\end{equation}
where $r^*$ is the tortoise coordinate
\cite{MTW}. For $\eta= +1$ ($\eta= -1$),
$T - \eta r^*$ is the retarded (advanced)
\eddfink\ time coordinate.
While $P_M$ was associated with the Killing time \cite{kuchar1}, our
prospective new momentum $\Pi_M$ is therefore associated with the retarded
or advanced \eddfink\ time.

Away from the shell, a calculation of the Poisson brackets shows that the
set $\left(M, \sfR, \Pi_M, \Pi_\sfR\right)$ is a candidate for a new
gravitational canonical chart. We need to find shell variables that
complete this set into a full canonical chart.

For the rest of this subsection, we assume that the shell
history does not lie on a horizon.
The special case where the shell straddles a horizon will
be discussed in subsection~\ref{subsec:trans-horizon}.

As a preliminary, suppose that the constraints
(\ref{metric-constr-eoms}) hold, and consider the regularity of
the variables. Away from the shell, the constraints
(\ref{metric-constr-eoms}) imply that $R'$ and $P_\Lambda$ are continuous,
and $F_\pm$ are thus both continuous. In the notation of
appendix~\ref{app:dist-eoms-gen}, the distributional content
(\ref{delta-constraints}) of the constraints at the shell can be written as
\begin{mathletters}
\label{delta-constraints2}
\begin{eqnarray}
0
&=&
\Delta F_{-\eta}
\ \ ,
\label{delta-constraints2-F}
\\
0
&=&
\zp + \Delta (\Lambda P_\Lambda)
\ \ .
\label{delta-constraints2-PL}
\end{eqnarray}
\end{mathletters}
{}From (\ref{delta-constraints2-F}) we see that $F_{-\eta}$ is
continuous at the shell. Equation (\ref{PiM-def}) then implies that $\Pi_M$
is  continuous, with the exception that it diverges on the horizon that is
parallel to the shell history. The first equality sign in~(\ref{PisfR-def}),
and the observation that the vacuum constraints are linear combinations of
$M'$ and $P_\sfR$ \cite{kuchar1}, imply that $\Pi_\sfR$ is vanishing
everywhere except possibly at the shell. The rightmost expression in
(\ref{PisfR-def}) shows that $\Pi_\sfR$ cannot contain a delta-function at
the shell, and $\Pi_\sfR$ is therefore everywhere vanishing. {}From now on,
we can therefore proceed assuming that $F_{-\eta}$ and $\Pi_\sfR$ are
continuous, and that their $r$-derivatives have at most finite
discontinuities at isolated values of~$r$. By~(\ref{PiM-def}), the same
will then hold for~$\Pi_M$, with the exception of the horizon where $\Pi_M$
diverges. This tightens the neighborhood of the classical solutions in
which the fields can take values, but it will not affect the critical
points of the action. The reason for this assumption is that it will make
the terms $\Pi_M {\dot{M}}$ and $\Pi_\sfR {\dot{\sfR}}$ in our new action
distributionally well defined.

We can now proceed to the Liouville forms. A direct computation yields
\begin{eqnarray}
P_\Lambda {\extder \Lambda}
+ P_R {\extder R}
+ M \extder{\Pi}_M - \Pi_\sfR {\extder \sfR}
&=&
- {\left( \eta R {\extder R} \ln \left| F_{-\eta} \right| \right)}'
\nonumber
\\
&&
+ \extder
\left[
{\eta R \Lambda \over 2}
\left( F^{-1}_{-\eta} - F_{-\eta} \right)
+ \eta R R' \ln \left| F_{-\eta} \right| \right]
\ \ .
\label{diff-liouville}
\end{eqnarray}
The variation $\extder$ affects the smoothness of
the gravitational variables in the same way as the time derivative in
subsection~\ref{subsec:local-eoms}. Away from the horizon parallel to the
shell history, on which $F_{-\eta}$ vanishes and ${\Pi}_M$ diverges, all
the terms in (\ref{diff-liouville}) are therefore distributionally well
defined: the terms on the left-hand side are at most discontinuous in~$r$,
while the terms on the right-hand side may contain at worst delta-functions
arising from~$(\extder R)'$. The status at the horizon on which $F_{-\eta}$
vanishes will be discussed below.

To obtain the difference in the prospective Liouville forms, we need to
integrate the relation (\ref{diff-liouville}) over~$r$. In an integral over
a finite interval in~$r$, the only subtlety arises from the horizon on
which $F_{-\eta}$ vanishes. On a classical solution with mass~$M$, it can
be shown from the embedding analysis of
appendix \ref{app:static-shell}
that
\begin{equation}
F_{-\eta} = {\Lambda_h \over 4 M} \, (r-r_h)
+ O \biglb( (r-r_h)^2 \bigrb)
\ \ ,
\label{Fminuseta-hor-fall}
\end{equation}
where the subscript $h$ indicates the values of the quantities at the
horizon on which $F_{-\eta}$ vanishes. Equations (\ref{Fminuseta-hor-fall})
and (\ref{PiM-def}) therefore show that, on the classical solution, the
integral of (\ref{diff-liouville}) across the horizon is well
defined in the principal value sense, just as in the corresponding
analysis of Ref. \cite{kuchar1}. To extend this argument off the classical
solutions, we note that when the constraints hold, $M(r)$ is constant in
$r$ across this horizon. As our action contains the constraints with their
associated Lagrange multiplies, we argue that $M(r)$ can be assumed smooth
at the  horizon in the relation~(\ref{diff-liouville}). We can then again
employ (\ref{Fminuseta-hor-fall}) and~(\ref{PiM-def}), and it is seen as
above that the integral of (\ref{diff-liouville}) across this horizon is
well defined in the principal value sense.

What needs more attention is the falloff in (\ref{diff-liouville}) at the
infinities. {}From (\ref{metricfall}), (\ref{trans}), and~(\ref{Pi-def}),
we have
\begin{mathletters}
\begin{eqnarray}
M(t,r)
&=&
M_{\pm}(t) + O^{\infty}({|r|}^{-\epsilon})
\ \ ,
\label{Mfall}
\\
\sfR(t,r)
&=&
|r| + O^{\infty}({|r|}^{-\epsilon})
\ \ ,
\label{sfRfall}
\\
\Pi_M(t,r)
&=&
\pm \eta \left( 1 + 2 M_\pm{|r|}^{-1} \right)
+ O^\infty \left( {|r|}^{-1-\epsilon} \right)
\ \ ,
\\
\Pi_\sfR(t,r)
&=&
O^\infty \left( {|r|}^{-1-\epsilon} \right)
\ \ .
\end{eqnarray}
\end{mathletters}
This means that the integrals of the third term on the left-hand side and
the total variation term on the right-hand side diverge as $r\to\pm\infty$.
The geometrical reason for this divergence is, as seen
from~(\ref{PiMprime}), that $\Pi_M$ is associated with a {\em null\/} time,
rather than an asymptotically Minkowski time.

The cure is to introduce convergence functions that provide
the necessary translation between asymptotically spacelike hypersurfaces and
asymptotically null hypersurfaces. To this end, let $g(M_+, M_-; r)$ be a
function that is smooth in $r$ and depends on our variables only through
$M_+$ and $M_-$ as indicated. Let $g$ have the falloff
\begin{equation}
g(M_+, M_-; r)
= \pm M_\pm^2{|r|}^{-1} + O^\infty ({|r|}^{-1-\epsilon})
\ \ .
\end{equation}
Adding $-\eta {\extder g}$ on both sides of (\ref{diff-liouville}) yields
now an equation whose both sides can be integrated in $r$ from $-\infty$
to~$\infty$. The substitution terms arising from the first
term on the right-hand side vanish, and we obtain
\begin{eqnarray}
\int_{-\infty}^\infty
(P_\Lambda {\extder \Lambda}
+ P_R {\extder R})
dr
&&=
\int_{-\infty}^\infty
(\Pi_\sfR {\extder \sfR} - M {\extder{\Pi}}_M +\eta {\extder g})
dr
\nonumber
\\
&&
+ \; \extder
\left\{
\int_{-\infty}^\infty
\left[
{\eta R \Lambda \over 2}
\left( F^{-1}_{-\eta} - F_{-\eta} \right)
+ \eta R R' \ln \left| F_{-\eta} \right|
-\eta g \right]
dr
\right\}
\ \ .
\label{int-diff-liouville}
\end{eqnarray}
All the terms in (\ref{int-diff-liouville}) are well defined, provided the
integral across the horizon on which $F_{-\eta}$ vanishes is interpreted in
the principal value sense. We therefore see that the set
$\left(M, \sfR, \zr, \Pi_M, \Pi_\sfR, \zp\right)$ provides a new
canonical chart on the phase space. Note that this canonical
transformation leaves the shell variables $(\zr,\zp)$ entirely invariant.
The geometrical meaning of the convergence function $g$ will 
be discussed in section~\ref{sec:reduction}.

What remains is to write the constraint terms in the action in terms of the
new variables. Consider first the constraints away from the shell.
A~straightforward rearrangement yields
\begin{equation}
NH + N^r H_r
=
N^\sfR \Pi_\sfR + {\tilde N} M'
\ \ ,
\label{smooth-new-constraints}
\end{equation}
where
\begin{mathletters}
\begin{eqnarray}
{\tilde N} &:=&
(\eta \Lambda N^r - N) F_{-\eta}^{-1}
\ \ ,
\\
N^\sfR &:=&
N^r R'
- N R^{-1} P_\Lambda
\ \ .
\end{eqnarray}
\end{mathletters}
Note that $N^\sfR$ is the same as in Ref.\ \cite{kuchar1}.
Both terms on the right-hand side of (\ref{smooth-new-constraints}) are
distributionally well defined.  Away from the shell, we can therefore
include the constraints in the action in the form shown on the right-hand
side of~(\ref{smooth-new-constraints}), with ${\tilde N}$ and $N^\sfR$ as
independent Lagrange multipliers. This constraint redefinition is mildly
singular on the horizon parallel to the shell history, owing to the
divergence of~$\Pi_M$; however, one can argue as in Ref.\
\cite{kuchar1} that the redefined constraints are equivalent to the old
ones by continuity. The falloff of the new multipliers is
\begin{mathletters}
\begin{eqnarray}
{\tilde N}
&=&
\mp N_\pm + O^\infty({|r|}^{-\epsilon})
\ \ ,
\\
N^\sfR
&=&
O^\infty({|r|}^{-\epsilon})
\ \ .
\end{eqnarray}
\end{mathletters}

To recover the delta-constraint~(\ref{delta-constraints2-F}), we observe
from (\ref{PisfR-def}) that  (\ref{delta-constraints2-F}) is equivalent to
$\Pi_\sfR$ not having a delta-contribution at the shell. We therefore
argue that including in the action the constraint term $-\int dt
\int_{-\infty}^\infty dr \, N^\sfR \Pi_\sfR$, with $N^\sfR$ an independent
Lagrange multiplier,  yields both the constraint $\Pi_\sfR=0$ away from the
shell and the delta-constraint (\ref{delta-constraints2-F}) at the
shell.\footnote{A subtlety in this argument is that $N^\sfR$ need not be
continuous at the shell, not even on the classical solutions. The product
$N^\sfR \Pi_\sfR$ would therefore not be distributionally well defined in
the event that $\Pi_\sfR$ did contain a delta-contribution at the shell.}

Finally, consider the delta-constraint~(\ref{delta-constraints2-PL}).
Using (\ref{PiM-def}) and~(\ref{delta-constraints2-F}), equation
(\ref{trans-M}) implies
\begin{equation}
\Delta M =
- {\Delta(\Lambda P_\Lambda) \over {\widehat{\Pi_M}} }
\ \ .
\label{DeltaM1}
\end{equation}
When (\ref{delta-constraints2-F}) holds,
(\ref{delta-constraints2-PL}) is thus equivalent to
\begin{equation}
\Delta M =
{\zp \over {\widehat{\Pi_M}}}
\ \ .
\label{DeltaM2}
\end{equation}
Including in the action the constraint term $-\int dt
\int_{-\infty}^\infty dr \, {\tilde N} [ M' - \zp \Pi_M^{-1}
\delta(r-\zr)]$, with ${\tilde N}$ an independent Lagrange multiplier,
therefore yields both the constraint $M'=0$ away from the shell and the
delta-constraint  (\ref{delta-constraints2-PL}) at the shell.

These considerations have led us to the action
\begin{eqnarray}
S
&=& \int dt
\left[ \zp {\dot\zr}
+ \int_{-\infty}^\infty
dr
\left(
\Pi_\sfR {\dot \sfR}
- M {\dot \Pi}_M
+ \eta {\dot g}
\right)
\right]
\nonumber
\\
&& - \int dt \;
\int_{-\infty}^\infty dr
\left\{
N^\sfR \Pi_\sfR
+ {\tilde N} [ M' - \zp \Pi_M^{-1} \delta(r-\zr) ]
\right\}
\nonumber
\\
&&-\int dt \, \left( N_+M_+ + N_- M_- \right)
\ \ .
\label{new-action-mpimdot}
\end{eqnarray}
Both the action and its variations are well defined. The Poisson bracket
algebra of the constraints clearly closes. Note that the convergence term
$\eta \dot{g}$ in no way contributes to the local variations of the action.

The Liouville term $- M {\dot \Pi}_M$ can be brought to a form in
which the time derivative is on~$M$, at the cost of introducing
another convergence term. Let $G(r)$ be a smooth function of $r$ only,
with the falloff
\begin{equation}
G(r) = \pm1 + O^\infty({|r|}^{-1-\epsilon})
\ \ .
\end{equation}
We then have
\begin{equation}
\eta {\dot g}
- M {\dot \Pi}_M
=
(\Pi_M - \eta G) \dot{M}  - \eta \dot{g}
+ {d\over dt}
\left(
\eta G M + 2 \eta g - M \Pi_M
\right)
\ \ .
\label{MPiM-int}
\end{equation}
All the terms in (\ref{MPiM-int}) are well defined, and each side can be
integrated in $r$ from $-\infty$ to~$\infty$. We arrive at the action
\begin{eqnarray}
S
&=& \int dt
\left\{ \zp {\dot\zr}
+ \int_{-\infty}^\infty
dr
\left[
\Pi_\sfR {\dot \sfR}
+ (\Pi_M - \eta G) \dot{M}  - \eta \dot{g}
\right]
\right\}
\nonumber
\\
&& - \int dt \;
\int_{-\infty}^\infty dr
\left\{
N^\sfR \Pi_\sfR
+ {\tilde N} [ M' - \zp \Pi_M^{-1} \delta(r-\zr)]
\right\}
\nonumber
\\
&&-\int dt \, \left( N_+M_+ + N_- M_- \right)
\ \ .
\label{new-action-pimmdot}
\end{eqnarray}
The geometrical meaning of the convergence terms
will become explicit in section~\ref{sec:reduction}.

\subsection{Shell on a horizon}
\label{subsec:trans-horizon}

In subsection \ref{subsec:trans-nonhorizon} we excluded the special
case where the shell straddles a horizon. We now discuss
how this special case can be included.

When the shell straddles a horizon, the zero of $F_{-\eta}$
occurs at the shell. The delta-constraints at the shell are given
by~(\ref{delta-constraints2}). When the equations of motion hold, the
masses on the two sides agree, and the embedding analysis of
appendix \ref{app:static-shell} shows that equation
(\ref{Fminuseta-hor-fall}) holds, now with
$\Lambda_h = \hat\Lambda$ and $r_h = \zr$. {}From (\ref{PisfR-def}) we see
that $\Pi_\sfR$ cannot contain a delta-function at the shell. We can
therefore again assume that $F_{-\eta}$ and $\Pi_\sfR$ are continuous, and
that their $r$-derivatives have at most finite discontinuities at isolated
values of~$r$.

The new feature in equation (\ref{diff-liouville}) is that the singularity
of $F_{-\eta}^{-1}$ and $\Pi_M$ now occurs at the shell. When the equations
of motion hold, we see from (\ref{PiM-def}) and (\ref{Fminuseta-hor-fall})
that integrating each side of (\ref{diff-liouville}) in $r$ across the
shell is well defined in the principal value sense, and we argue as
above that this conclusion can be extended away from the classical
solutions provided the constraints are understood to hold. We argue
similarly that the left-hand side of (\ref{diff-liouville}) contains no
delta-contributions at the shell, and it is therefore justified to
interpret the integral of (\ref{diff-liouville}) over $r$ as the principal
value. Convergence at the infinities is accomplished as above, and the
substitution terms from the total $r$-derivative on the right-hand side
of (\ref{diff-liouville}) vanish. We therefore again arrive
at~(\ref{int-diff-liouville}). Equations (\ref{DeltaM1}) and
(\ref{DeltaM2}) remain valid, with the understanding
${\left({\widehat{\Pi_M}}\right)}^{-1}=0$, and the delta-constraints 
can be taken in the action as before. To justify the manipulations leading
to the action~(\ref{new-action-pimmdot}), we again appeal to the constraints
to argue that $M$ can be regarded as smooth in $r$ at the shell, and
that $\dot{M}$ then does not contain a delta-function at the shell.

We therefore see that the actions (\ref{new-action-mpimdot}) and
(\ref{new-action-pimmdot}) remain valid in a suitable limiting sense also
for a horizon-straddling shell.

\section{Reduction}
\label{sec:reduction}

In this section we eliminate the constraints and find the dynamics in the
reduced phase space. We shall continue to treat the cases $\eta=\pm1$
separately, and we denote the corresponding two components of the reduced
phase space by~$\gtildeta$. We first assume, in subsection
\ref{subsec:reduction-not-hor}, that the shell history does not lie on a
horizon, and we then include the horizon-straddling shell as a limiting
case in subsection \ref{subsec:reduction-on-hor}. Finally, in subsection
\ref{subsec:reduction-no-shell}, we attach the right-moving and
left-moving test shell limits to the respective components of the reduced
phase space as regular boundaries.

It will be useful in the reduction to assume a more definite form for 
the convergence function~$g$. {}From now on, we take
\begin{mathletters}
\label{g-special}
\begin{equation}
g(M_+, M_-; r)
=
M_+^2 g_+(r) + M_-^2 g_-(r)
\ \ ,
\end{equation}
where $g_\pm(r)$ are smooth functions of $r$ only, with the falloff
\begin{equation}
g_\pm(r) = \pm {|r|}^{-1} \theta(\pm r)
+ O^\infty({|r|}^{-1-\epsilon})
\ \ ,
\label{g-special-fall}
\end{equation}
\end{mathletters}
where $\theta$ denotes the step function.

\subsection{Shell not on a horizon}
\label{subsec:reduction-not-hor}

In this subsection we assume that the shell history does not lie on a
horizon.

Solving the constraint $\Pi_\sfR=0$ implies that $\sfR$ and $\Pi_\sfR$
simply drop out of the action. To solve the remaining constraint,
\begin{equation}
\Pi_M M' - \zp
\delta(r-\zr)=0
\ \ ,
\label{red-constr-pim}
\end{equation}
we write
\begin{equation}
M =
\bmplus\theta(r-\zr)
+
\bmminus \theta(\zr-r)
\ \ ,
\label{M-sol}
\end{equation}
where $\bmplusminus(t)$ are regarded as independent variables. We then
have
\begin{equation}
\dot{M}(r) =
{\dot{\bm}}_+ \theta(r-\zr)
+ {\dot{\bm}}_- \theta(\zr-r)
+ (\bmminus - \bmplus) \dot{\zr} \delta(r-\zr)
\ \ ,
\label{dotM-red}
\end{equation}
and the constraint (\ref{red-constr-pim}) implies
\begin{equation}
\zp = (\bmplus - \bmminus) {\widehat{\Pi_M}}
\ \ .
\label{zp-sol-red}
\end{equation}
Note that as the shell history does not lie on a horizon, each of the two
factors on the right-hand side of (\ref{zp-sol-red}) is nonvanishing.

Using~(\ref{g-special-fall}), (\ref{dotM-red}), and~(\ref{zp-sol-red}),
we find
\begin{eqnarray}
\zp {\dot\zr} + \int_{-\infty}^{\infty} dr
\left[
(\Pi_M - \eta G) \dot{M}  - \eta \dot{g}
\right]
&=& \bpplus \dot{\bm}_+
+ \bpminus \dot{\bm}_-
\nonumber
\\
&&+
{d\over dt}
\left[
\eta(\bmplus-\bmminus) \int_0^\zr G \, dr
\right]
\ \ ,
\label{lioville-reduction}
\end{eqnarray}
where
\begin{mathletters}
\label{bmpplusminus}
\begin{eqnarray}
\bpplus &:=& \int_{-\infty}^\infty dr
\left[
\Pi_M \theta(r-\zr) - \eta G \theta(r) - 2\eta \bmplus g_+
\right]
\ \ ,
\label{bmpplus}
\\
\bpminus &:=& \int_{-\infty}^\infty dr
\left[
\Pi_M \theta(\zr-r) - \eta G \theta(-r) - 2\eta \bmminus g_-
\right]
\ \ .
\label{bmpminus}
\end{eqnarray}
\end{mathletters}
The singularity of $\Pi_M(r)$ occurs in precisely one of the two integrals
in~(\ref{bmpplusminus}), and the integral over this singularity is
interpreted in the principal value sense. Substituting
(\ref{lioville-reduction}) into~(\ref{new-action-pimmdot}), and dropping
the integral of a total derivative, we obtain the reduced action
\begin{equation}
S = \int dt
\left(
\bpplus {\dot{\bf m}}_+
+
\bpminus {\dot{\bf m}}_-
- N_+ \bmplus - N_- \bmminus \right)
\ \ .
\label{red-action1}
\end{equation}
This shows that the set $(\bmplus, \bmminus, \bpplus, \bpminus)$ provides
local canonical coordinates on~$\gtildeta$.
The equations of motion derived from the action (\ref{red-action1}) read
\begin{mathletters}
\begin{eqnarray}
{\dot{\bf m}}_\pm  &=& 0
\ \ ,
\\
{\dot{\bf p}}_\pm &=& - N_\pm
\label{dotbpplusminus}
\ \ .
\end{eqnarray}
\end{mathletters}

The emergence of $\bmplusminus$ as two coordinates on $\gtildeta$ is not
surprising: on a classical solution, $\bmplusminus$ are the two
Schwarzschild masses, and these masses together with $\eta$ completely
determine the four-dimensional spacetime. To understand the geometrical
meaning of~$\bpplusminus$, we recall from (\ref{PiMprime}) that without the
convergence terms proportional to $G$ and~$g_\pm$, the integrals in
(\ref{bmpplusminus}) would give the \eddfink\ time differences between the
shell and the infinities on the constant $t$ hypersurface. As the constant
$t$ hypersurface extends to the spacelike infinities, such null-time
differences would be infinite. The role of the convergence terms in
(\ref{bmpplusminus}) is to absorb the infinities: one can think of the
convergence terms as associating to the constant $t$ hypersurface a
hypersurface that is asymptotically {\em null\/} as $r\to\pm\infty$. For
$\eta=-1$, this associated hypersurface extends from the left-hand-side
$\scriplus$ to the right-hand-side~$\scriminus$ (points $A$ and $B$ in
figure~\ref{fig:nonhorizon}); for $\eta=1$, the situation is the reverse.
Thus, $\bpplus$ is the \eddfink\ time difference between the shell and the
right-hand-side infinity of the associated asymptotically null
hypersurface, and $-\bpminus$ is the \eddfink\ time difference between the
shell and the left-hand-side infinity of the associated asymptotically null
hypersurface. The equations of motion (\ref{dotbpplusminus}) show that the
time evolution of $\bpplusminus$ only arises from the evolution of the
constant $t$ hypersurfaces at the infinities. Thus, in this local canonical
chart on~$\gtildeta$, the information about the shell motion is encoded in
equations~(\ref{dotbpplusminus}).

It should be emphasized that the degrees of freedom present in
$\bpplusminus$ are invariant under the isometries of the spacetime. Killing
time translations on the spacetime move both the shell history and the
constant $t$ hypersurface: in particular, they move the two asymptotic ends
of the constant $t$ hypersurface, and hence the asymptotic ends of the
associated asymptotically null hypersurface. However, the \eddfink\ time
{\em differences\/} that constitute the momenta are invariant under Killing
time translations.

As we have assumed that the shell history does not lie on a horizon, the
coordinates $(\bmplus, \bmminus, \bpplus, \bpminus)$ do not form a global
chart on~$\gtildeta$. Instead, these coordinates provide two disjoint local
canonical charts, covering two disconnected sets in~$\gtildeta$: one for
$0<\bmminus<\bmplus$ and the other for $0<\bmplus<\bmminus$, with
unrestricted values of $\bpplusminus$ in each chart. These coordinates
cannot be extended to $\bmplus=\bmminus$. The reason is that when
$\bmplus=\bmminus$, the shell history lies on a horizon, the singularity in
$\Pi_M$ is at $r=\zr$, and the first term under each integral in
(\ref{bmpplusminus}) makes both $\bpplus$ and $\bpminus$ divergent. We
shall address the special case $\bmplus=\bmminus$  and the global structure
of $\gtildeta$ in subsection~\ref{subsec:reduction-on-hor}.

It will be useful to introduce on $\gtildeta$ 
another set of local canonical
coordinates, $(\bmbar, \bmtilde, \bpbar, \bptilde)$, 
by the transformation
\begin{mathletters}
\label{plusminus-to-tildebars}
\begin{eqnarray}
\bmbar &=& \casehalf ( \bmplus + \bmminus )
\ \ ,
\\
\bmtilde &=& \casehalf ( \bmplus - \bmminus )
\ \ ,
\\
\bpbar &=& \bpplus + \bpminus
\ \ ,
\label{bpbar-def}
\\
\bptilde &=& \bpplus - \bpminus
\label{bptilde-def}
\ \ .
\end{eqnarray}
\end{mathletters}
The inverse transformation is
\begin{mathletters}
\begin{eqnarray}
\bmplusminus &=& \bmbar \pm \bmtilde
\ \ ,
\\
\bpplusminus &=& \casehalf ( \bpbar \pm \bptilde )
\ \ .
\end{eqnarray}
\end{mathletters}
{}From (\ref{bmpplusminus}), (\ref{bpbar-def}), and~(\ref{bptilde-def}), we
see that $\bpbar$ contains the information about the asymptotic ends of the
constant $t$ hypersurface, whereas the information about the location of
the shell with respect to the infinities is encoded in~$\bptilde$. We can
therefore loosely regard the pair $(\bmbar,\bpbar)$ as describing the vacuum
spacetime dynamics, and the pair $(\bmtilde,\bptilde)$ as describing the
shell. While not literal, this view will be helpful for understanding the
global structure of $\gtildeta$ in subsections \ref{subsec:reduction-on-hor}
and~\ref{subsec:reduction-no-shell}.

As defined by the transformation~(\ref{plusminus-to-tildebars}), the
coordinates $(\bmbar, \bmtilde, \bpbar, \bptilde)$ provide two disjoint
local canonical charts that cover on $\gtildeta$ the same two disconnected
sets as the coordinates $(\bmplus, \bmminus, \bpplus, \bpminus)$. The
ranges of the variables in these two charts are respectively
$0<\bmtilde<\bmbar$ and $0<-\bmtilde<\bmbar$, each with unrestricted
$\bpbar$ and~$\bptilde$. The coordinates $(\bmbar,
\bmtilde, \bpbar, \bptilde)$ cannot, however, be extended to $\bmtilde=0$.
While $\bpbar$ remains finite for a horizon-straddling shell, it is seen
from (\ref{bmpplusminus}) that $\bptilde$ must diverge.

\subsection{Shell on a horizon}
\label{subsec:reduction-on-hor}

We now wish to find on $\gtildeta$ coordinates that extend to the
horizon-straddling shell. We shall first rely on the spacetime picture to
identify the geometrical information that the coordinates must carry in
this limit. We then construct on $\gtildeta$ a global canonical chart that
contains this information.

Consider the spacetime of figure~\ref{fig:nonhorizon}. The shell is
left-moving, corresponding to $\eta=-1$, and the shell history lies
in the future of the left-going horizon, corresponding to
$\bmplus>\bmminus$. The points $A$ and $B$ indicate the ends of the
asymptotically null hypersurface that is associated to the hypersurface of
constant~$t$. $\bpplus$~is the difference in the \eddfink\ time
between points $p_1$ and~$q_1$, and $\bpminus$ is the difference in the
\eddfink\ time between points $q_2$ and~$p_2$.

In this spacetime, let $\gamma_1$ be the radial null geodesic connecting
$p_1$ to~$q_1$, let $\gamma_2$ the radial null geodesic connecting $q_2$
to~$p_2$, and let $\gamma_3$ be the radial null geodesic connecting $p_1$
to~$p_2$.  Let $\lambda_i$ ($i=1,2,3$) be the affine parameters on these
geodesics, each normalized to have the range $[0,1]$. $\lambda_2$~and
$\lambda_3$ increase toward the future. $\lambda_1$~increases toward the
future if $q_1$ is in the future of $p_1$ as shown in the figure,
corresponding to $\bpplus<0$, and it increases towards the past if $q_1$ is
in the past of~$p_1$, corresponding to $\bpplus>0$. In the special case
$\bpplus=0$, $q_1$~and $p_1$ coincide, and $\gamma_1$
degenerates to a point. We now define the quantities $Q_\pm$ by
\begin{mathletters}
\label{Q12-def}
\begin{eqnarray}
Q_+ &:=&
\left.
{(\partial/\partial \lambda_1)}_a {(\partial/\partial \lambda_3)}^a
{\vphantom{A^A}} \right|_{p_1}
\ \ ,
\\
\noalign{\smallskip}
Q_- &:=&
\left.
{(\partial/\partial \lambda_2)}_a {(\partial/\partial \lambda_3)}^a
{\vphantom{A^A}} \right|_{p_2}
\ \ .
\end{eqnarray}
\end{mathletters}

Similarly, consider a spacetime in which $\eta=-1$ but the shell history
lies in the past of the left-going horizon, corresponding to
$\bmplus<\bmminus$. In this spacetime, the counterparts of points $p_1$ and
$p_2$ are below the left-going horizon, but the three null geodesics
$\gamma_i$ ($i=1,2,3$) can be defined as above, the only modification being
that the potentially degenerate one is now~$\gamma_2$. In this spacetime,
we again define $Q_\pm$ by~(\ref{Q12-def}).

A straightforward calculation yields
\begin{mathletters}
\label{Qpm-ab-bel}
\begin{eqnarray}
Q_+ &=&
-8 \bmplus
\left[ \bmminus - \bmplus
+ |\bmplus-\bmminus|
\exp\left(-\frac{\bpplus}{4\bmplus}\right)
\right]
\ \ ,
\\
\noalign{\smallskip}
Q_- &=&
-8 \bmminus
\left[ \bmplus - \bmminus
+ |\bmplus-\bmminus|
\exp\left(\frac{\bpminus}{4\bmminus}\right)
\right]
\ \ ,
\end{eqnarray}
\end{mathletters}
valid both for $\bmplus>\bmminus$ and $\bmplus<\bmminus$.
The canonical coordinates $(\bmplus, \bmminus, \bpplus, \bpminus)$ can
therefore be replaced by the noncanonical coordinates $(\bmplus, \bmminus,
Q_+,Q_-)$, with $Q_-<0$ for $\bmplus>\bmminus$ and $Q_+<0$ for
$\bmplus<\bmminus$.

The crucial observation is now that~$Q_\pm$, as defined
in~(\ref{Q12-def}), remain well defined also for the spacetimes shown in
figure~\ref{fig:horizon}, in which $\bmplus=\bmminus$ and the shell
history lies on a common horizon. In these spacetimes, $Q_\pm$ can each
take arbitrary negative values. As the soldering along the horizon is
affine, $Q_\pm$ precisely encode the coordinate-invariant information
about the relative loci of the points $p_1$, $p_2$, $A$, and~$B$ (or,
equivalently, the points $p_1$, $p_2$, $q_1$, and~$q_2$). This means that
the set $(\bmplus,\bmminus, Q_+,Q_-)$ provides a global, noncanonical
chart
on~${\tilde\Gamma}_-$.
The domain is $Q_-<0$ for $\bmplus>\bmminus$, $Q_+<0$ for
$\bmplus<\bmminus$, and $Q_\pm<0$ for $\bmplus=\bmminus$.

In the above construction we have taken $\eta=-1$. It is clear that
an entirely analogous discussion carries through for
$\eta=1$, with straightforward changes in formulas~(\ref{Qpm-ab-bel}), and
yielding a global, noncanonical chart on on~${\tilde\Gamma}_+$.

To find a global {\em canonical\/} chart on~$\gtildeta$,
consider the transformation
\begin{mathletters}
\label{bpi-def}
\begin{eqnarray}
\bpibar &:=& \bpbar + 8 \eta \bmtilde
\left( \ln|\bmtilde/\bmbar| -1 \right)
\ \ ,
\\
\bpitilde &:=& \bptilde + 8 \eta \bmbar
\left( \ln|\bmtilde/\bmbar| +1 \right)
\ \ .
\end{eqnarray}
\end{mathletters}
Equations (\ref{bpi-def}) clearly define a canonical transformation from
$(\bmbar,\bmtilde,\bpbar,\bptilde)$ to
$(\bmbar,\bmtilde,\bpibar,\bpitilde)$ individually in the domains
$0<\bmtilde<\bmbar$ and $0<-\bmtilde<\bmbar$. It is straightforward to
verify that the chart $(\bmbar,\bmtilde,\bpibar,\bpitilde)$ becomes global
on $\gtildeta$ when extended to $\bmtilde=0$ with unrestricted values of
$\bpibar$ and~$\bpitilde$. For $\eta=-1$, in particular, (\ref{Qpm-ab-bel})
shows that $Q_\pm$ can be written as
\begin{mathletters}
\begin{eqnarray}
Q_+ &=&
-16 (\bmbar + \bmtilde)
\left\{
\bmbar \exp \!
\left[
- \left(
\frac{\bmbar - \bmtilde}{\bmbar + \bmtilde}
\right)
-
\frac{\bpibar + \bpitilde}{8(\bmbar + \bmtilde)}
\right]
-\bmtilde
\right\}
\ \ ,
\\
\noalign{\smallskip}
Q_- &=&
-16 (\bmbar - \bmtilde)
\left\{
\bmbar \exp \!
\left[
- \left(
\frac{\bmbar + \bmtilde}{\bmbar - \bmtilde}
\right)
+
\frac{\bpibar - \bpitilde}{8(\bmbar - \bmtilde)}
\right]
+\bmtilde
\right\}
\ \ ,
\end{eqnarray}
\end{mathletters}
from which the regularity of the $\bmtilde\to0$ limit is manifest.

We have thus shown that the set $(\bmbar,\bmtilde,\bpibar,\bpitilde)$
provides a global canonical chart on~$\gtildeta$. The domain of
the variables is $|\bmtilde|<\bmbar$, with $\bpibar$ and $\bpitilde$ taking
all real values.
We therefore have $\gtildeta \simeq \BbbR^4$.
The Hamiltonian reads
\begin{equation}
\bh = (N_+ + N_-) \bmbar + (N_+ - N_-) \bmtilde
\ \ .
\end{equation}
The values of $\bmbar$ and $\bmtilde$ are constants of motion, whereas the
equations of motion for $\bpibar$ and $\bpitilde$ show that the
evolution of $Q_\pm$ only arises from the evolution of the constant $t$
hypersurface at the two spacelike infinities. This means that the
information about the shell dynamics is contained in the momentum equations
of motion both for $\bmtilde\ne0$ (as was already seen in subsection
\ref{subsec:reduction-not-hor}) as well as for $\bmtilde=0$.

\subsection{Test shell limit}
\label{subsec:reduction-no-shell}

We have so far assumed that the unreduced shell momentum $\zp$ is
nonvanishing. We saw that this assumption is compatible with the dynamics,
and that it divides the reduced phase space into the two disconnected
sectors~$\gtildeta$, labeled by $\eta=\sign(\zp)$. As the unreduced bulk
action (\ref{S-ham}) is not differentiable in $\zp$ at $\zp=0$, it is not
clear whether ${\tilde\Gamma}_+$ and ${\tilde\Gamma}_-$ are joinable to
each other in any smooth sense. Our reduction formalism is not well suited
to examining this issue: the canonical transformation of
section~\ref{sec:transformation} was tailored to  the null hypersurfaces
separately for $\eta=\pm1$.

We can, however, address the limit $\zp\to0$ {\em individually\/}
in ${\tilde\Gamma}_+$ and ${\tilde\Gamma}_-$. As the shell stress-energy
tensor (\ref{shell-Tab}) vanishes for $\zp\to0$, this is the
limit of a test shell that traverses the spacetime without affecting it
gravitationally. We shall now show that one can attach the right-moving and
left-moving test shell limits respectively to ${\tilde\Gamma}_+$ and
${\tilde\Gamma}_-$ as smooth boundaries with topology~$\BbbR^3$.

When the test shell history does not lie on a horizon, the situation is
straightforward. We can start with the coordinates
$(\bmplus, \bmminus, \bpplus, \bpminus)$, separately for
$0<\bmminus<\bmplus$ and $0<\bmplus<\bmminus$, and simply take the limit
$\bmplus=\bmminus$ with $\bpplusminus$ remaining finite. {}From the
geometrical interpretation of $\bpplusminus$ it is seen that this attaches
to $\gtildeta$ those test shell configurations in which the
test shell does not straddle a horizon. The locus of the test shell
history is determined by $\bpplusminus$ exactly as in
subsection~\ref{subsec:reduction-not-hor}.

Including a horizon-straddling test shell is more intricate. In the
global
chart $(\bmbar,\bmtilde,\bpibar,\bpitilde)$, the limit of a test shell on a
horizon is achieved by setting first $\bmtilde=0$ and then taking
$\eta\bpitilde\to-\infty$ while keeping $\bpibar$ finite. On the other
hand, the limit of a test shell off the horizon requires taking
simultaneously $\bmtilde\to0$ and $\eta\bpitilde\to-\infty$
so that $\bpbar$ and $\bptilde$ remain finite.
What we need is a new canonical chart in which both of these limits are
brought to finite values of the coordinates.

To this end, let
\begin{mathletters}
\label{bd-chart}
\begin{eqnarray}
x &:=&
\exp \!
\left(\frac{\eta\bpitilde}{8\bmbar}\right)
\ \ ,
\\
p_x &:=&
- 8 \eta \bmtilde\bmbar
\exp \!
\left(- \frac{\eta\bpitilde}{8\bmbar}\right)
\ \ ,
\\
\bPibar &:=& \bpibar- \frac{\bpitilde \bmtilde}{\bmbar}
\ \ .
\end{eqnarray}
\end{mathletters}
As
\begin{equation}
\bPibar \extder \bmbar
+ p_x \extder x
=
\bpibar \extder \bmbar
+ \bpitilde \extder \bmtilde
- \extder (\bpitilde\bmtilde)
\ \ ,
\end{equation}
equations (\ref{bd-chart}) define a canonical transformation from the chart
$(\bmbar,\bmtilde,\bpibar,\bpitilde)$ to the chart $(\bmbar,x,\bPibar,p_x)$.
The new canonical chart is global: the range is $\bmbar>0$ and
$x>0$, with unrestricted $\bPibar$ and~$p_x$. The qualitative location of
the shell history is governed by the sign of~$p_x$: $p_x>0$ ($p_x<0$) yields
a shell in the future (past) of the horizon that is parallel to
the shell history, while $p_x=0$ yields a shell history on the horizon. It
is now easily seen that in this chart the test shell limit is $x\to0$, with
the other coordinates remaining at finite values. A horizon-straddling test
shell is recovered with $p_x=0$, whereas $p_x>0$ ($p_x<0$) gives a test
shell in the future (past) of the horizon that the test shell does not
cross. Clearly, the test shell limit constitutes a smooth boundary of
$\gtildeta$ with topology~$\BbbR^3$.

\section{Hamiltonian for the shell radius in stationary exterior
coordinates}
\label{sec:gauges}

While the charts on $\gtildeta$ introduced in section \ref{sec:reduction}
are well adapted to the geometry of the spacetime, they contain the
information about the shell motion in a nontransparent manner. In this
section we introduce on $\gtildeta$ a local canonical chart that
describes more directly the motion of the shell in the spacetime geometry.
A chart of this kind is of particular physical interest if one wishes
to quantize the system as a model of black hole radiation with back reaction
\cite{kraus-wilczek1,kraus-wilczek2,vakkuri-kraus}.

The physical situation we have in mind is a static observer who
scrutinizes the shell motion from an asymptotically flat infinity. 
For definiteness, we take this infinity to be the right-hand-side
one. We set $N_+=1$, so that the coordinate time $t$ coincides with the
observer's proper time. To incorporate the observer's ignorance of what is
happening at the left-hand-side infinity, we set $N_-=0$.

We further assume that the shell history reaches a future or past null
infinity on the right-hand side. The Penrose diagram for $\eta=-1$ is
therefore as in figure~\ref{fig:nonhorizon}, and the Penrose diagram for
$\eta=1$ is the time inverse. In particular, we have $\bmminus<\bmplus$, and
we are in the region of $\gtildeta$ covered by the chart
$(\bmplus,\bmminus,\bpplus,\bpminus)$ with $0<\bmminus<\bmplus$.

Consider thus the chart $(\bmplus,\bmminus,\bpplus,\bpminus)$ with
$0<\bmminus<\bmplus$.{}From section \ref{sec:reduction} we recall that the
pair $(\bmplus,\bpplus)$ only carries information about the geometry right
of the shell, and the pair $(\bmminus,\bpminus)$ only carries information
about the geometry left of the shell. To describe the motion of the shell
as seen from the right-hand-side infinity, we can therefore leave the
pair $(\bmminus,\bpminus)$ intact and seek a canonical transformation that
replaces $(\bmplus,\bpplus)$ by a new pair. 

To specify the new pair, we choose in the Kruskal geometry
right of the shell a stationary coordinate system that conforms to the
falloff (\ref{metricfall}) with $N_+=1$. As $R(r)$ is then an increasing
function, we can assume $R(r)=r$ without loss of generality. Let $\rhat$
stand for the shell curvature radius in these coordinates:
$\rhat(t) := R\biglb(\zr(t)\bigrb) = \zr(t)$. We now seek a momentum $\phat$
such that there is a canonical transformation from the pair
$(\bmplus,\bpplus)$ to the pair $(\rhat,\phat)$. Writing the
transformation as $\bpplus = \bpplus(\rhat, \bmplus)$ and
$\phat = \phat(\rhat, \bmplus)$, the canonicality criterion reads 
\begin{equation}
{\partial [\phat(\rhat, \bmplus)] \over \partial \bmplus}
= -
{\partial [\bpplus(\rhat, \bmplus)] \over \partial \rhat}
\ \ .
\label{can-crit}
\end{equation}
Substituting $\bpplus$ from
(\ref{bmpplus}) to the right-hand side of (\ref{can-crit}) yields
\begin{equation}
{\partial [\phat(\rhat, \bmplus)] \over \partial \bmplus}
=
\Pi_M(\rhat, \bmplus)
\ \ ,
\label{phatvpi}
\end{equation}
where $\Pi_M(r,\bmplus)$ is determined by the choice of the stationary
coordinate system. Note that the convergence functions $G(r)$ and $g_+(r)$
have not entered~(\ref{phatvpi}). Solving the differential equation
(\ref{phatvpi}) for $\phat(\rhat,\bmplus)$ yields the desired canonical
transformation, and inverting this solution gives $\bmplus(\rhat,\phat)$ as
a function in the new canonical chart. The action reads
\begin{equation}
S = \int dt \,
[\bpminus {\dot\bmminus}
+ \phat \dot{\rhat}
- \bmplus(\rhat,\phat)]
\ \ .
\end{equation}
The shell stress-energy tensor in the new chart can be found
using~(\ref{ell-def}),  (\ref{shell-Tab}) and~(\ref{zp-sol-red}).

As explicit examples, we now present the shell Hamiltonians
$\bmplus(\rhat,\phat)$ in four different stationary coordinate
systems. We arrived at the first three coordinate systems by seeking
a simple functional form for $\bmplus(\rhat,\phat)$. The fourth coordinate
system is the spatially flat one used in Ref.\ \cite{kraus-wilczek1}.

\subsection{Polynomial gauge}

As a first example, we consider coordinates in which the metric reads
\begin{mathletters}
\label{polygauge-metric}
\begin{eqnarray}
&&R
=
r
\ \ ,
\\
&&\Lambda^2
=
N^{-2} =
1 + 2\bmplus/r + {(2\bmplus/r)}^2 + {(2\bmplus/r)}^3
\ \ ,
\\
&&\Lambda^2 N^r
=
- \eta
{(2\bmplus/r)}^2
\ \ .
\end{eqnarray}
\end{mathletters}
With $r>0$, these coordinates cover half of Kruskal manifold in the
appropriate manner, and the falloff (\ref{metricfall}) is satisfied with
$\epsilon=1$. The relation to the curvature
coordinates is
\begin{mathletters}
\begin{eqnarray}
R &=& r
\ \ ,
\\
T &=& t + 2\eta\bmplus \ln |1 - 2\bmplus/r|
\ \ .
\end{eqnarray}
\end{mathletters}
We find
\begin{equation}
\Pi_M(r, \bmplus) = \eta ( 1 + 2\bmplus/r)
\ \ .
\end{equation}
Solving (\ref{phatvpi}) with a convenient choice for the integration
constant, we obtain
\begin{equation}
\bmplus(\rhat,\phat)
=
\sqrt{\eta \phat \rhat} - \casehalf \rhat
\ \ .
\label{polygauge-ham}
\end{equation}

The equation of motion derived from the Hamiltonian (\ref{polygauge-ham})
can be integrated as $\eta t = \rhat + 2\bmplus \ln (
\rhat/\bmplus ) + \hbox{constant}$. It is easily verified that this is the
correct equation for a null geodesic in the
metric~(\ref{polygauge-metric}).

\subsection{Exponential gauge}

Consider next coordinates in which the metric reads
\begin{mathletters}
\begin{eqnarray}
&&R
=
r
\ \ ,
\\
&&\Lambda^2
=
N^{-2} =
\exp(2\bmplus/r) [2 - (1 - 2\bmplus/r)\exp(2\bmplus/r)]
\ \ ,
\\
&&\Lambda^2 N^r
=
- \eta
\left[ 1 - (1 - 2\bmplus/r)\exp(2\bmplus/r) \right]
\ \ .
\end{eqnarray}
\end{mathletters}
With $r>0$, these coordinates cover half of Kruskal manifold in the
appropriate manner, and the falloff (\ref{metricfall}) is satisfied with
$\epsilon=1$. The relation to the curvature
coordinates is
\begin{mathletters}
\begin{eqnarray}
R &=& r
\ \ ,
\\
T &=& t + \eta \int^r
\left[ {(1 - 2\bmplus/r')}^{-1} - \exp(2\bmplus/r') \right]
dr'
\ \ .
\end{eqnarray}
\end{mathletters}
We find
\begin{equation}
\Pi_M(r, \bmplus) = \eta \exp(2\bmplus/r)
\ \ .
\end{equation}
Solving (\ref{phatvpi}) with a convenient choice for the integration
constant, we obtain
\begin{equation}
\bmplus(\rhat,\phat)
=
\casehalf \rhat \ln(2\eta \phat/\rhat)
\ \ .
\end{equation}
The equation of motion can be solved implicitly in terms of the exponential
integral function.

\subsection{\eddfink--type gauge}

Consider next coordinates in which the metric reads
\begin{mathletters}
\label{eddfinkgauge-metric}
\begin{eqnarray}
&&R
=
r
\ \ ,
\\
&&\Lambda^2
=
N^{-2} =
1 + 2\bmplus/r
\ \ ,
\\
&&\Lambda^2 N^r
=
-
2 \eta \bmplus/r
\ \ .
\end{eqnarray}
\end{mathletters}
With $r>0$, these coordinates cover half of Kruskal manifold in the
appropriate manner. The relation to the curvature
coordinates is
\begin{mathletters}
\begin{eqnarray}
R &=& r
\ \ ,
\\
T &=& t + 2 \eta \bmplus \ln |r/(2\bmplus) - 1|
\ \ .
\end{eqnarray}
\end{mathletters}
We recognize these coordinates as simply related to the \eddfink\
coordinates \cite{MTW}: $t-\eta r$ is the retarded (advanced) \eddfink\ time
for $\eta=1$ ($\eta=-1$). In terms of the tortoise
coordinate $r^* :=  r + 2 \bmplus \ln [r/(2\bmplus) - 1]$,
we have $t-\eta r = T - \eta r^*$.

There is a minor technical issue in that the coordinates
(\ref{eddfinkgauge-metric}) do not obey the falloff~(\ref{metricfall}):
we have $P_\Lambda = - 2 \eta \bmplus {(1 + 2\bmplus/r)}^{-1/2}$, which
violates~(\ref{PLambdafall}). We therefore take the coordinates
(\ref{eddfinkgauge-metric}) to hold for $r<\Rcut$, where $\Rcut$ is a large
parameter, and smoothly deform them to a faster falloff for $r>\Rcut$. As
equation (\ref{phatvpi}) is local in~$\rhat$, the form of the canonical
transformation for $r<\Rcut$ is independent of that for $r>\Rcut$. In the
end, we can either leave the Hamiltonian unspecified for $r>\Rcut$, or
argue that one can take the limit $\Rcut\to\infty$ in the sense of some
suitable renormalization in the parameters of the canonical transformation. 

Proceeding in this way, we find
\begin{equation}
\Pi_M(r, \bmplus) = \eta
\ \ ,
\end{equation}
and, with a convenient choice of the integration constant,
\begin{equation}
\bmplus(\rhat,\phat)
= \eta \phat
\ \ .
\label{efgauge-ham}
\end{equation}
The Hamiltonian (\ref{efgauge-ham}) clearly correctly reproduces the fact
that the \eddfink\ time $t-\eta r$ is constant on the shell history.

\subsection{Spatially flat gauge}
\label{subsec:spatflat-gauge}

As the last example, we consider coordinates in which the metric reads
\begin{mathletters}
\label{spatflatgauge-metric}
\begin{eqnarray}
&&R
=
r
\ \ ,
\\
&&\Lambda = N
=
1
\ \ ,
\\
&&N^r
=
- \eta \sqrt{2\bmplus/r}
\ \ .
\end{eqnarray}
\end{mathletters}
With $r>0$, these coordinates cover half of Kruskal manifold in the
appropriate manner. The relation to the curvature
coordinates is
\begin{mathletters}
\begin{eqnarray}
R &=& r
\ \ ,
\\
T &=& t + 2 \eta \left( \sqrt{2 \bmplus r} + \bmplus
\ln
\left|
{1 - \sqrt{2\bmplus/r} \over 1 + \sqrt{2\bmplus/r}}
\right|
\right)
\ \ .
\end{eqnarray}
\end{mathletters}
We recognize these coordinates as the spatially flat coordinates
\cite{painleve,gullstrand,israel-flat}, recently employed in the study of
Hawking radiation with back reaction in Ref.\ \cite{kraus-wilczek1}.

There is again a minor technical issue in that the coordinates
(\ref{spatflatgauge-metric}) do not obey the falloff~(\ref{metricfall}). A
Hamiltonian falloff analysis compatible with these coordinates has been
discussed in the metric variables in Ref.\ \cite{FriLoWi}. Here, however, we
shall simply argue in terms of a cutoff parameter $\Rcut$ as above.  We have
\begin{equation}
\Pi_M(r, \bmplus) =
{\eta \over 1 + \sqrt{2 \bmplus/r}}
\ \ ,
\end{equation}
and solving (\ref{phatvpi}) with a suitable integration constant yields
$\bmplus(\rhat,\phat)$ implicitly as the solution to
\begin{equation}
\eta \phat = \sqrt{2\bmplus \rhat} - \rhat \ln \left( 1 +
\sqrt{2\bmplus/\rhat} \right)
\ \ .
\label{bmplus-implicit}
\end{equation}

In order to make a connection to the work in Ref.\ \cite{kraus-wilczek1},
we define
\begin{mathletters}
\label{KW-momenta}
\begin{eqnarray}
p_c
&:=&
\phat - \eta \left[
\sqrt{2\bmminus \rhat} - \rhat \ln \left( 1 +
\sqrt{2\bmminus/\rhat} \right)
\right]
\ \ ,
\label{KW-momenta-pc}
\\
\bp_c
&:=&
\bpminus - \eta \left[
\rhat - 2 \sqrt{2\bmminus\rhat}
+ 4 \bmminus \ln \left( 1 +
\sqrt{\rhat/(2\bmminus)} \right)
\right]
\ \ .
\end{eqnarray}
\end{mathletters}
Equations (\ref{KW-momenta}) define a canonical transformation from the
chart $(\rhat,\bmminus,\phat,\bpminus)$ to the new canonical chart
$(\rhat,\bmminus,p_c,\bp_c)$. The Hamiltonian $\bmplus(\rhat,\bmminus,p_c)$
in the new chart is obtained (in implicit form) by eliminating $\phat$ from
(\ref{bmplus-implicit}) and~(\ref{KW-momenta-pc}). As the value of
$\bmminus$ is a constant of motion, the system can be partially reduced
by regarding $\bmminus$ as a prescribed constant. The term $\int dt \,
\bp_c {\dot\bmminus}$ then drops from the action,
and we obtain
\begin{equation}
S = \int dt \,
[p_c \dot{\rhat}
- \bmplus(\rhat,\bmminus,p_c)]
\ \ .
\end{equation}
For $\eta=-1$, this is the action derived in Refs.\
\cite{kraus-wilczek1,FriLoWi} by different methods. For $\eta=1$, it is
not. The reason is that the coordinates (\ref{spatflatgauge-metric}) are
the ingoing spatially flat coordinates for $\eta=-1$ and the outgoing
spatially flat coordinates for $\eta=1$, thus covering all of the spacetime
right of the shell in each case, whereas Ref.\
\cite{kraus-wilczek1} was physically motivated to use the ingoing
spatially flat coordinates irrespectively the direction of the shell motion.
It would be straightforward to repeat the above analysis with the sign of
$\eta$ in (\ref{spatflatgauge-metric}) reversed, recovering the result of
Refs.\ \cite{kraus-wilczek1,FriLoWi} for $\eta=1$. Note, however, that with
the sign of $\eta$ in (\ref{spatflatgauge-metric}) reversed, the coordinates
do not cover the part of the shell history that lies inside the horizon.

\section{Parametrization clocks at the infinities}
\label{sec:para-times}

In the previous sections we fixed the evolution of the spatial
hypersurfaces at the spacelike infinities by taking $N_\pm$ to be prescribed
functions of~$t$. In this section we free this evolution by making the
replacement
\cite{kuchar1}
\begin{equation}
N_\pm = \pm{\dot\tau}_\pm
\label{N-vs-tau}
\end{equation}
in the boundary term in the actions (\ref{total-metric-action})
and~(\ref{new-action-pimmdot}). The variations of $N$ become then
unrestricted at $r\to\pm\infty$, but varying the action with respect
to $\tau_\pm$ yields the relations (\ref{N-vs-tau}) as equations of motion.
The new variables $\tau_\pm$ are the proper times measured by static
standard clocks at the respective infinities, with the convention that
$\tau_+$ increases toward the future and $\tau_-$ increases toward the
past. 

In the absence of a shell, it was shown in Ref.\ \cite{kuchar1} that the
action containing $\tau_\pm$ as independent variables can be brought to a
canonical form in which the unconstrained degrees of freedom and the pure
gauge degrees of freedom are entirely decoupled. We now outline the
analogous result in the presence of the null shell. For brevity, we shall
refrain from explicitly spelling out the smoothness properties of the
various emerging phase space functions.

We start from the action~(\ref{new-action-pimmdot}), with $g$ given
by~(\ref{g-special}), and we make in the boundary term the
replacement~(\ref{N-vs-tau}). The resulting action is a sum of
two decoupled parts: a Hamiltonian action $S_\sfR$ consisting of the terms
that contain the pair $(\sfR,\Pi_\sfR)$, and the remainder~$S_0$. We only
need to consider~$S_0$. As in section~\ref{sec:transformation}, we assume
first that the shell history does not lie on a horizon, and relax this
assumption at the end.

Under the time integral in~$S_0$, the terms homogeneous in the time
derivatives are
\begin{equation}
\Theta :=  \zp {\dot\zr}
- M_+ {\dot\tau}_+ + M_- {\dot\tau}_-
+
\int_{-\infty}^{\infty} dr
\left[
(\Pi_M - \eta G) \dot{M}  - \eta \dot{g}
\right]
\ \ .
\end{equation}
We pass from the noncanonical chart $\biglb(\zr,M(r),\zp,\Pi_M(r);
\tau_+,\tau_-\bigrb)$ to the new chart
$\biglb(\mbar, \zr, \Gamma(r), \pbar, \zp,\Pi_\Gamma(r)\bigrb)$,
defined by
\begin{mathletters}
\begin{eqnarray}
\Gamma(r) &:=& M'(r)
\ \ ,
\label{Gamma}
\\
\Pi_\Gamma(r)
&:=&
\casehalf (\tau_+ + \tau_-)
+ \int_{-\infty}^\infty dr' \,
\biggl\{
\casehalf [\Pi_M(r') - \eta G(r')]
\left[
\theta(r'-r) - \theta(r-r')
\right]
\nonumber
\\
&& \qquad\qquad\qquad\qquad\qquad\quad
- \eta M_+ g_+(r') + \eta M_- g_-(r')
\biggr\}
\ \ ,
\label{PiGamma}
\\
\mbar &:=& \casehalf (M_+ + M_-)
\ \ ,
\label{mbardef}
\\
\pbar &:=&
\tau_+ - \tau_-
+ \int_{-\infty}^\infty dr \,
\biggl\{
[\Pi_M(r) - \eta G(r)]
- 2\eta M_+ g_+(r) - 2\eta M_- g_-(r)
\biggr\}
\ \ .
\end{eqnarray}
\end{mathletters}
The falloff is
\begin{mathletters}
\label{GammaPiGammafall}
\begin{eqnarray}
\Gamma(t,r)
&=&
O^{\infty}({|r|}^{-1-\epsilon})
\ \ ,
\\
\Pi_\Gamma(t,r)
&=&
-2 \eta M_\pm \ln|r/M_\pm|
+ O^{\infty}({|r|}^{0})
\ \ .
\end{eqnarray}
\end{mathletters}
By techniques similar to those in Ref.\ \cite{kuchar1}, we find
\begin{equation}
\Theta =  \zp {\dot\zr}
+ \pbar \, \dot{\mbar}
+ \int_{-\infty}^\infty dr \, \Pi_\Gamma(r) \dot{\Gamma}(r)
+ {d\over dt}
\left(
M_+\tau_+ + M_-\tau_-
\right)
\ \ .
\end{equation}
The chart $\biglb(\mbar, \zr, \Gamma(r), \pbar, \zp,\Pi_\Gamma(r)\bigrb)$ is
therefore canonical. Dropping the integral of a total derivative, the
action reads
\begin{eqnarray}
S_0 &=& \int dt \,
\left(
\zp {\dot\zr}
+ \pbar \, \dot{\mbar}
+ \int_{-\infty}^\infty dr \,
\Pi_\Gamma \dot{\Gamma}
\right)
\nonumber
\\
&&
- \int dt \int_{-\infty}^\infty dr \,
\tilde{N}
\biggl[
\Gamma
- \zp {(\eta G - \Pi'_\Gamma)}^{-1} \delta(r-\zr)
\biggr]
\ \ .
\label{Snought-can1}
\end{eqnarray}
To express the constraint in the new chart, we have used equation
(\ref{Gamma}) and the relation $\Pi_M = \eta G - \Pi'_\Gamma$,
which follows by differentiating~(\ref{PiGamma}).

The action~(\ref{Snought-can1}) is canonical, but the constraint couples
the variables in a nontransparent way. To decouple the degrees of freedom,
we pass to the chart
$\biglb(\mbar,\mtilde,\tilde\Gamma(r),
\pbar,\ptilde,\Pi_{\tilde\Gamma}(r)\bigrb)$,
defined by
\begin{mathletters}
\begin{eqnarray}
\tilde\Gamma(r)
&:=&
\Gamma(r)
- \zp {\left[\eta G(r) - \Pi'_{\Gamma}(r) \right]}^{-1} \delta(r-\zr)
\ \ ,
\label{tildeGamma}
\\
\Pi_{\tilde\Gamma}(r)
&:=&
\Pi_\Gamma(r)
\ \ ,
\\
\mtilde
&:=&
\casehalf
\zp {\left(\eta \hat{G} - \widehat{\Pi'_{\Gamma}} \right)}^{-1}
\ \ ,
\label{mtildedef}
\\
\ptilde
&:=&
2 \widehat{\Pi_\Gamma} - 2 \eta \int_0^\zr dr \, G(r)
\ \ .
\end{eqnarray}
\end{mathletters}
The falloff of $\tilde\Gamma$ and $\Pi_{\tilde\Gamma}$ is clearly the same
as that of~$\Gamma$ and $\Pi_\Gamma$, given in~(\ref{GammaPiGammafall}).
Using the analogue of relation (\ref{total-der-Rhat}) for~$\Pi_\Gamma$, we
find
\begin{equation}
\zp {\dot\zr} - \ptilde \dot{\mtilde}
+ \int_{-\infty}^\infty dr \,
\left(
\Pi_\Gamma \dot{\Gamma} - \Pi_{\tilde\Gamma} \dot{\tilde{\Gamma}}
\right)
= {d\over dt}
\left[
2\eta \mtilde \int_0^\zr dr \, G(r)
\right]
\ \ .
\end{equation}
The chart $\biglb(\mbar,\mtilde,\tilde\Gamma(r),
\pbar,\ptilde,\Pi_{\tilde\Gamma}(r)\bigrb)$ is therefore canonical.
Dropping the integral of a total derivative, the action reads
\begin{eqnarray}
S_0 &=& \int dt \,
\left(
\pbar \, \dot{\mbar}
+ \ptilde \dot{\mtilde}
+ \int_{-\infty}^\infty dr \,
\Pi_{\tilde\Gamma} \dot{\tilde\Gamma}
\right)
\nonumber
\\
&&
- \int dt \int_{-\infty}^\infty dr \,
\tilde{N} \tilde{\Gamma}
\ \ .
\end{eqnarray}
The unconstrained canonical degrees of freedom,
$(\mbar,\mtilde,\pbar,\ptilde)$, have now become 
decoupled from the pure gauge degrees of freedom.

To put the action in a more transparent form, we write
\begin{mathletters}
\begin{eqnarray}
V(r)
&:=&
\Pi_{\tilde\Gamma}(r) + \casehalf \pbar
- \eta \int_0^r dr' \, G(r')
\nonumber
\\
&=&
\tau_+
+ \int_{-\infty}^\infty dr'
\left[
\Pi_M(r') \theta(r'-r)
- \eta G(r') \theta(r')
- 2 \eta M_+ g_+(r')
\right]
\ \ ,
\label{Vrdef}
\\
\Pi_V(r)
&:=&
- \tilde{\Gamma}(r)
\ \ .
\end{eqnarray}
\end{mathletters}
The falloff is
\begin{mathletters}
\label{VPVfall}
\begin{eqnarray}
V(t,r)
&=&
- \eta |r| -2 \eta M_\pm \ln|r/M_\pm|
+ O^{\infty}({|r|}^{0})
\ \ ,
\\
\Pi_V(t,r)
&=&
O^{\infty}({|r|}^{-1-\epsilon})
\ \ .
\end{eqnarray}
\end{mathletters}
As
\begin{equation}
\int_{-\infty}^\infty dr
\left(
\Pi_{\tilde\Gamma} \dot{\tilde{\Gamma}}
- \Pi_V \dot{V}
\right)
=
- \casehalf \dot{\pbar}
\int_{-\infty}^\infty dr \,
\Pi_V
+ {d\over dt}
\int_{-\infty}^\infty dr \, \Pi_{\tilde\Gamma} {\tilde{\Gamma}}
\ \ ,
\label{dotpbarPiV}
\end{equation}
the transformation to the chart
$\biglb(\mbar,\mtilde,V(r),\pbar,\ptilde,\Pi_V(r)\bigrb)$ is not canonical
as it stands. However, it becomes canonical after the first
term on the right-hand side of (\ref{dotpbarPiV}) is absorbed into the
constraint term by writing
\begin{equation}
N^V :=
- \tilde{N} + \casehalf \dot{\pbar}
\end{equation}
and regarding $N^V$ as a new Lagrange multiplier. As the
equations of motion imply $\dot{\pbar}=0$, the falloff of $N^V$ is
\begin{equation}
N^V =
\pm N_\pm + O^\infty({|r|}^{-\epsilon})
\ \ .
\end{equation}
Dropping the integral of a total derivative, and
including~$S_\sfR$, we finally obtain the action
\begin{eqnarray}
S &=& \int dt
\left(
\pbar \, \dot{\mbar}
+ \ptilde \dot{\mtilde}
\right)
\nonumber
\\
&&
+ \int dt
\int_{-\infty}^\infty dr
\left(
\Pi_V \dot{V}
+\Pi_\sfR {\dot \sfR}
- N^V \Pi_V
- N^\sfR \Pi_\sfR
\right)
\ \ .
\label{fheis-action}
\end{eqnarray}

All the variables in the action (\ref{fheis-action}) have a
transparent geometrical meaning.
{}From~(\ref{Gamma}),
(\ref{mbardef}),
(\ref{tildeGamma}),
and~(\ref{mtildedef}), we see that $\mbar$ and $\mtilde$ are respectively
equal to the variables $\bmbar$ and $\bmtilde$ introduced in
section~\ref{sec:reduction}. Similarly, using (\ref{N-vs-tau}), we
see that $\pbar$ and $\ptilde$ can be interpreted as
the time-independent initial values of the variables $\bpbar$ and $\bptilde$
introduced in section~\ref{sec:reduction}. As for the pure gauge degrees of
freedom, $\sfR$ is the curvature radius, and equation (\ref{Vrdef}) shows
that $V$ is the \eddfink\ time. The action (\ref{fheis-action}) therefore
provides a natural generalization of the vacuum action given in Eq.~(149)
of Ref.\ \cite{kuchar1}.

We have here assumed that the shell history does not lie on a horizon. This
assumption can be relaxed, in a suitable limiting sense, by performing on
the coordinates $(\mbar,\mtilde,\pbar,\ptilde)$ transformations analogous
to those given for the coordinates
$(\bmbar,\bmtilde,\bpbar,\bptilde)$ in section~\ref{sec:reduction}.

\section{$\BbbR^3$ spatial topology}
\label{sec:flat-int}

In this section we consider the canonical transformation and Hamiltonian
reduction for spatial topology~$\BbbR^3$. For concreteness, we take the
evolution of the spatial hypersurfaces at the single spacelike infinity  to
be prescribed as in section~\ref{sec:metric}.  It will be seen that the
reduced phase space consists of two disconnected components, one for an
expanding shell and the other for a collapsing shell. Each component has
the topology~$\BbbR^2$.

We start from the action principle. In the bulk action~(\ref{S-ham}), we
take $0<r<\infty$, with the falloff (\ref{metricfall}) as $r\to\infty$. 
As $r\to0$, we introduce the falloff
\begin{mathletters}
\label{s-metricfall}
\begin{eqnarray}
\Lambda(t,r)
&=&
\Lambda_0 + O(r^2)
\ \ ,
\label{s-Lambdafall}
\\
R(t,r)
&=&
R_1 r + O(r^3)
\ \ ,
\label{s-Rfall}
\\
P_\Lambda(t,r)
&=&
P_{\Lambda_2} r^2 + O(r^4)
\ \ ,
\label{s-PLambdafall}
\\
P_R(t,r) &=&
P_{R_1} r + O(r^3)
\ \ ,
\label{s-PRfall}
\\
N(t,r)
&=&
N_0 +
O(r^2)
\ \ ,
\label{s-Nfall}
\\
N^r(t,r)
&=&
N^r_1 r + O(r^3)
\ \ ,
\label{s-Nrfall}
\end{eqnarray}
\end{mathletters}
where $\Lambda_0>0$, $R_1>0$, $P_{\Lambda_2}$, $P_{R_1}$, $N_0>0$, and
$N^r_1$ are functions of $t$ only. It is straightforward to verify that the
falloff (\ref{s-metricfall}) is consistent with the constraints and
preserved by the time evolution. By (\ref{trans-M}) and~(\ref{F-def}),
the falloff (\ref{s-metricfall}) implies that the mass left of the shell
must vanish when the equations of motion hold: $r=0$ is then just the
coordinate singularity at the center of hyperspherical coordinates in flat
space. The classical solutions therefore describe a shell with a flat
interior, and the spatial topology is~$\BbbR^3$. The action appropriate for
fixing $N_+$ is
\begin{equation}
S = S_\Sigma - \int dt \, N_+ M_+
\ \ ,
\end{equation}
where $S_\Sigma$ is given by (\ref{S-ham}) with $0<r<\infty$.

The canonical transformation of section \ref{sec:transformation} goes
through with the obvious changes. The new action is as
in~(\ref{new-action-mpimdot}), except that the integral is from $r=0$ to
$r=\infty$ and the term $N_- M_-$ is missing. $G(r)$~and $g(M_+;r)$ are
smooth in $r$ and have the same behavior as 
$r\to+\infty$ as in section~\ref{sec:transformation}.
The falloff of the new fields as $r\to0$ can be found
from~(\ref{s-metricfall}); for example, we have
\begin{equation}
\tilde{N}(t,r) = - N_0 \Lambda_0 R_1^{-1}
+ O(r)
\ \ .
\end{equation}
All the new fields remain regular as $r\to0$. In particular, $M(r)$
tends to zero as $r\to0$.

The Hamiltonian reduction proceeds as in section~\ref{sec:reduction}, with
the simplification that the interior mass vanishes.
The reduced phase space consists again of two disconnected components,
denoted now by ${\tilde\Gamma}^{E}_\eta$.
We take $g(M_+;r)$ to be as in (\ref{g-special}) with $M_-=0$, and we solve
the constraint (\ref{red-constr-pim}) as in (\ref{M-sol}) and
(\ref{zp-sol-red}) with
$\bmminus=0$. Note that as $\Pi_M$ has the same sign as~$\zp$, equation
(\ref{zp-sol-red}) implies $\bmplus>0$. We find
\begin{equation}
\zp {\dot\zr} + \int_0^{\infty} dr
\left[
(\Pi_M - \eta G) \dot{M}  - \eta \dot{g}
\right]
= \bpplus \dot{\bm}_+
+
{d\over dt}
\left(
\eta \bmplus \int_0^\zr G \, dr
\right)
\ \ ,
\end{equation}
where
\begin{equation}
\bpplus := \int_0^\infty dr
\left[
\Pi_M \theta(r-\zr) - \eta G  - 2\eta \bmplus g_+
\right]
\ \ .
\label{bpplus-flat}
\end{equation}
Substituting this in the action and dropping the integral of a total
derivative, we obtain the reduced action
\begin{equation}
S = \int dt
\left(
\bpplus {\dot\bmplus}
- N_+ \bmplus \right)
\ \ .
\end{equation}
Thus, the pair $(\bmplus,\bpplus)$ provides a canonical chart on
${\tilde\Gamma}^{E}_\eta$. As $\Pi_M$ does not have singularities,
the definition (\ref{bpplus-flat}) is always good: the chart is
global, and the topology of ${\tilde\Gamma}^{E}_\eta$ is~$\BbbR^2$.
The test shell limit can be attached as a smooth boundary with topology
$\BbbR$ at $\bmplus=0$.

The information about the shell motion is again encoded in the evolution
of~$\bpplus$. Charts that describe the shell motion in the exterior
geometry more transparently can be constructed as in
section~\ref{sec:gauges}.

\section{Summary and discussion}
\label{sec:discussion}

In this paper we have analyzed the Hamiltonian structure of spherically
symmetric Einstein gravity coupled to an infinitesimally thin null-dust
shell. We formulated the theory under Kruskal-like boundary conditions,
prescribing the evolution of the spatial hypersurfaces at the two
spacelike infinities. We adopted smoothness conditions that made the
variational equations distributionally well defined, and equivalent to the
Einstein equations for this system.

We then simplified the constraints by a Kucha\v{r}-type canonical
transformation and performed the Hamiltonian reduction. It was seen that
the reduced phase space consists of two disconnected copies of~$\BbbR^4$,
one for a right-moving shell and the other for a left-moving shell. We found
on each component a global canonical chart in which the configuration
variables are the Schwarzschild masses on the two sides of the shell,
leaving the shell dynamics indirectly encoded in the conjugate momenta.
Excluding the special case of a shell straddling a horizon, we found a local
canonical chart in which the configuration variables are the shell
curvature radius and the interior mass,  in an arbitrarily specifiable
stationary coordinate system exterior to the shell. In particular,
performing a partial reduction and fixing the interior mass to be a
prescribed constant, we reproduced a previously known shell Hamiltonian in
the spatially flat gauge outside the shell.

We also cast into canonical form the theory in which the evolution at the
infinities is freed by introducing parametrization clocks. We found on the
unreduced phase space a global canonical chart in which the physical
degrees of freedom and the pure gauge degrees of freedom are completely
decoupled, and we identified the pure gauge configuration variables in this
chart as the \eddfink\ time and the curvature radius. Finally, we adapted
the analysis to the spatial topology~$\BbbR^3$, which has just one
infinity, and for which the spacetime inside the shell is flat. Expectedly,
the reduced phase space for this spatial topology turned out to consist of
two disconnected copies of~$\BbbR^2$, one for an expanding shell and the
other for a collapsing shell.

In addition to the Kruskal spatial topology $S^2\times\BbbR$ and the
Euclidean spatial topology~$\BbbR^3$, yet another spatial topology of
interest would be that of the $\RPthree$ geon \cite{topocen},
$\RPthree\setminus${}$\{$a point at infinity$\}$. As the reduced phase space
of the vacuum theory with the $\RPthree$ geon topology has dimension two
\cite{LW2}, one expects that the reduced phase space with a null shell would
have dimension four. Indeed, this is the conclusion reached under a
technically slightly different but qualitatively similar falloff in Ref.\
\cite{FriLoWi}, by first performing a Hamiltonian reduction for a
massive dust shell and then taking the zero rest mass limit. It does not
seem straightforward to adapt the canonical transformation of section
\ref{sec:transformation} to $\RPthree$ geon topology, however. An
$\RPthree$-geon--type spacetime with a null shell can be mapped to a
Kruskal-type spacetime with two null shells, but these two shells must be
moving in opposite directions; our canonical transformation, on the other
hand, was adapted to only one direction of the shell motion at a time.

A similar issue arises if one wishes to include more than one null-dust
shell. One expects our canonical transformations to generalize readily to
the case when all the shells are moving in the same direction. Shells
moving in different directions would, however, seem to require new methods.

Several steps in our analysis relied crucially on the fact that the shell
is null. This issue appears first in the consistency of the ADM
equations of motion in section~\ref{sec:metric}. In a fixed background
geometry, the equations obtained by varying the action (\ref{S-ham}) with
respect to the shell variables must, by construction, be equivalent to the
geodesic equation for the shell. In our dynamical equations
(\ref{metric-dyn-eoms}), the pair consisting of (\ref{rhat-eom})
and~(\ref{rhatmom-eom}), if interpreted individually on
each side of the shell,
must therefore be equivalent to the null geodesic equation. The reason why
the potentially ambiguous equation (\ref{rhatmom-eom}) turns out to be
unambiguous is precisely that the junction is along a null hypersurface,
and this hypersurface is geodesic
in the geometries on both sides of the junction.

Next, the fact that the shell history is null led us to the
\eddfink\ time as a spacetime function that is sufficiently smooth to
provide an acceptable momentum conjugate to~$M(r)$. Finally, in
section~\ref{sec:gauges}, the null character of the shell history
made it possible to leave the interior canonical pair $(\bmminus,\bpminus)$
untouched in the canonical transformation from $(\bmplus,\bpplus)$ to
$(\rhat,\phat)$. This is because the null history, when viewed from the
exterior geometry, does not contain information about the interior mass,
beyond the statement that $\bmminus<\bmplus$.

These special properties of a null shell suggest that our analysis may not
be immediately generalizable to timelike shells. For example, for a dust
shell with a positive rest mass, already the consistency of the ADM
equations of motion fails under our smoothness assumptions: the variational
equations corresponding to  (\ref{metric-constr-eoms}) and
(\ref{metric-dyn-eoms}) only become consistent if the right-hand side in
the counterpart of (\ref{rhatmom-eom}) is by hand interpreted as its
average over the two sides of the shell \cite{FriLoWi}. However, new
avenues may open if one relaxes the assumption that the variations of the
geometry and matter be independent. Recent progress in this direction has
been made by H\'aj\'{\i}\v{c}ek and Kijowski
\cite{haji-kijo1,haji-gensshell,haji-kijo2}.

The work in this paper has been purely classical. One may, however, hope 
that our canonical charts on the reduced phase space  will prove useful for 
quantizing the system. In the spatially flat gauge outside the shell, the
quantization of the shell variables with fixed interior mass was introduced
as a model for Hawking radiation with back-reaction in Ref.\
\cite{kraus-wilczek1}, and the same approach was applied to related black
holes in Refs.\ \cite{kraus-wilczek2,vakkuri-kraus}. Our results provide the
tools for a similar analysis in an arbitrarily specifiable stationary gauge
outside the shell. Whether this freedom in the gauge choice can be utilized
to a physically interesting end remains to be seen.
One may also wish to explore quantizations based on the global canonical
charts in which the dynamics is simpler but the spacetime picture more
hidden. This might shed light on the analogous question of quantizing in a
dynamically simple but geometrically nontransparent canonical chart in the
context of a two-dimensional dilatonic gravity theory coupled to scalar
fields \cite{KRV}. We leave these questions
subject to future work.

\acknowledgments
We would like to thank
Tevian Dray,
Petr H\'aj\'{\i}\v{c}ek,
Jerzy Kijowski,
Karel Kucha\v{r},
Charlie Misner,
Eric Poisson,
and 
Stephen Winters
for discussions.
This work was supported in part by NSF grants
PHY94-08910, 
PHY94-21849,
and PHY95-07740.

\appendix
\section{Hamiltonian equations of motion at the shell}
\label{app:dist-eoms-gen}

In this appendix we isolate the independent information that the
Hamiltonian equations of motion, (\ref{metric-constr-eoms})
and~(\ref{metric-dyn-eoms}), contain at the shell. It will be shown in
appendices \ref{app:nonstatic-shell} and \ref{app:static-shell} that
when this information is combined to Einstein's equations away from the
shell, we unambiguously recover the correct junction conditions for general
relativity coupled to a null-dust shell.

To begin, we note that equations (\ref{metric-constr-eoms}) and all save the
last one of equations (\ref{metric-dyn-eoms}) have an unambiguous
distributional interpretation.  The constraint equations
(\ref{metric-constr-eoms}) contain explicit delta-functions in $r$ from the
matter contribution and implicit delta-functions in $R''$ and~$P_\Lambda'$.
The right-hand sides of (\ref{Lambda-eom}) and (\ref{R-eom}) contain at
worst finite discontinuities, and the right-hand sides of
(\ref{PLambda-eom}) and (\ref{PR-eom}) contain at worst delta-functions;
this is consistent with the left-hand sides of
(\ref{Lambda-eom})--(\ref{PR-eom}), recalling that the loci of
nonsmoothness in $\Lambda$, $R$, $P_\Lambda$ and $P_R$ may evolve smoothly
in~$t$. Wherever explicit or implicit delta-functions appear, they are
multiplied by continuous functions of~$r$. The only potentially troublesome
equation is therefore~(\ref{rhatmom-eom}): the right-hand side is a
combination of spatial derivatives evaluated at the shell, but our
assumptions allow these derivatives to be discontinuous.

If $f$ stands for any of our metric functions that may be discontinuous
at the shell, we define 
\begin{equation}
\Delta f := \lim_{\epsilon\to 0_+}
\left[ f(\zr + \epsilon)
- f(\zr - \epsilon) \right]
\ \ .
\label{Delta-def}
\end{equation}
The delta-contributions to $f'$ and $\dot{f}$ at the shell can then be
written respectively as $(\Delta f) \delta(r-\zr)$ and 
$-{\dot\zr} (\Delta f) \delta(r-\zr)$. With this notation, the constraint
equations (\ref{metric-constr-eoms}) at the shell read 
\begin{mathletters}
\label{delta-constraints}
\begin{eqnarray}
&&\Delta R'  = - \eta \zp
/
{\hat R}
\ \ ,
\label{delta-ham}
\\
\noalign{\smallskip}
&&\Delta P_\Lambda = - \zp / {\hat\Lambda}
\ \ ,
\label{delta-mom}
\end{eqnarray}
\end{mathletters}
and the delta-contributions in the dynamical equations (\ref{PLambda-eom})
and (\ref{PR-eom}) at the shell read
\begin{mathletters}
\label{delta-mom-eqs}
\begin{eqnarray}
- {\dot \zr} \Delta P_\Lambda
&=&
{\eta \zp {\hat N} \over {\hat \Lambda}^2}
+ {\widehat{N^r}} \Delta P_\Lambda
\ \ ,
\label{delta-momLambda-eq}
\\
- {\dot \zr} \Delta P_R
&=&
- { {\hat N} \Delta R' + {\hat R} \Delta N'
\over
{\hat \Lambda}}
+ {\widehat{N^r}} \Delta P_R
\ \ .
\label{delta-momR-eq}
\end{eqnarray}
\end{mathletters}
The full set of equations at the shell therefore consists of
(\ref{rhat-eom}), (\ref{rhatmom-eom}), (\ref{delta-constraints}),
and~(\ref{delta-mom-eqs}). Of these, all except (\ref{rhatmom-eom}) are
manifestly well defined.

Two of the six equations are easily seen to be redundant. First, inserting
$\Delta P_\Lambda$ from (\ref{delta-mom}) into (\ref{delta-momLambda-eq})
yields an equation that is proportional to (\ref{rhat-eom}) by the factor
$\zp/{\hat\Lambda}$. Equation (\ref{delta-momLambda-eq})  can therefore be
dropped. Second, by continuity of the metric, we observe that $\hat R (t)
= R\biglb(t,\zr(t)\bigrb)$ is well defined for all~$t$, and so is its total
time derivative, given by
\begin{equation}
\dot{\hat R} =
[\dot{R} + \dot{\zr} R']^{\outwidehat}
\ \ .
\label{total-der-Rhat}
\end{equation}
The individual terms on the right-hand side of (\ref{total-der-Rhat}) are
not continuous at the shell, but the left-hand side shows that the sum
must be, and we obtain
\begin{equation}
\Delta {\dot R} = - \dot{\zr} \Delta R'
\ \ .
\label{deltaprimedot}
\end{equation}
An entirely similar reasoning leads to counterparts of
(\ref{deltaprimedot}) with $R$ replaced by any metric function that is
continuous in~$r$.  Using (\ref{deltaprimedot}) and~(\ref{rhat-eom}),
equation (\ref{PLambda}) gives $\Delta P_\Lambda = \eta {\hat R} (\Delta
R') /{\hat \Lambda}$. This shows that the two equations in
(\ref{delta-constraints}) are equivalent, and we can
drop~(\ref{delta-mom}).

To simplify~(\ref{delta-momR-eq}), we evaluate $\Delta P_R$ from
(\ref{PLambda}) and eliminate $\dot{R}$ and $\dot{\Lambda}$ using
(\ref{deltaprimedot}) and its counterpart for~$\Lambda$.
Using~(\ref{rhat-eom}), the result can be arranged to read
\begin{equation}
0 = \Delta \left[ (v_a v^a)' \right]
\ \ ,
\label{delta-momR-eq-prime}
\end{equation}
where the vector field $v^a$ is defined by
\begin{mathletters}
\label{vee-def}
\begin{eqnarray}
v^t &=& 1
\ \ ,
\\
v^r &=& \dot\zr
\ \ ,
\end{eqnarray}
\end{mathletters}
both at the shell and away from the shell. At the shell, $v^a$ coincides
with the shell history tangent vector $\ell^a$~(\ref{ell-def}), by virtue
of the equation of motion~(\ref{rhat-eom}). Through standard manipulations,
(\ref{delta-momR-eq-prime}) can be brought to the form
\begin{equation}
0 = \Delta \left[ v_b v^a \nabla_a (\partial_r)^b \right]
\ \ ,
\label{delta-momR-eq-cov}
\end{equation}
where $\nabla_a$ is the spacetime covariant derivative.

The information in the Hamiltonian equations of motion at the shell is
therefore captured by the set consisting of~(\ref{rhat-eom}),
(\ref{rhatmom-eom}), (\ref{delta-ham}), and~(\ref{delta-momR-eq-cov}).

In appendices \ref{app:nonstatic-shell} and \ref{app:static-shell} we
combine these four equations to the fact that away from the shell,
equations (\ref{metric-constr-eoms}) and (\ref{metric-dyn-eoms}) are
equivalent to Einstein's equations and thus make the geometry locally
Schwarzschild. As noted in subsection \ref{subsec:local-eoms}, equation
(\ref{rhat-eom}) implies that the shell history is null. We therefore only
need to examine two qualitatively different cases, according to whether or
not the shell history lies on a horizon.  The results, derived respectively
in appendices \ref{app:nonstatic-shell} and~\ref{app:static-shell}, are
summarized here in the following two paragraphs:

When the shell history does not lie on a horizon, the continuity of $R$
across the shell completely determines the geometry. Equation
(\ref{delta-momR-eq-cov}) reduces to an identity, and the combination
of derivatives on the right-hand side of (\ref{rhatmom-eom}) is continuous
at the shell. Equation (\ref{rhatmom-eom}) becomes then well
defined. With $\zp$ given by~(\ref{delta-ham}),
equation (\ref{rhatmom-eom}) reduces to an identity.

When the shell history lies on a horizon, the masses on the two
sides must agree. Equation (\ref{delta-momR-eq-cov}) now implies
that the soldering along the junction is affine. The combination of
derivatives on the right-hand side of (\ref{rhatmom-eom}) is then
continuous at the shell, and equation (\ref{rhatmom-eom}) becomes well
defined. With $\zp$ given by~(\ref{delta-ham}), equation
(\ref{rhatmom-eom}) reduces to an identity.

\section{Equations of motion for a nonstatic shell}
\label{app:nonstatic-shell}

In this appendix we verify the claims in the penultimate paragraph of
appendix~\ref{app:dist-eoms-gen}. For concreteness, and without loss of
generality, we may assume that the shell history lies right of the horizon
that the shell does not cross. The geometry is then as in figure
\ref{fig:nonhorizon} for $\eta=-1$, and its time inverse for
$\eta=1$.

On each side of the shell, we introduce the \eddfink\
coordinates,
\begin{mathletters}
\label{EF-metric}
\begin{eqnarray}
ds^2 &=& - F dV^2 - 2 \eta dV dR  + R^2 d\Omega^2
\ \ ,
\\
F &=& 1 - 2M/R
\ \ ,
\label{EF-metric-F}
\end{eqnarray}
\end{mathletters}
where $M$ is the Schwarzschild mass. To avoid cluttering the
notation, we suppress indices that would distinguish the coordinate patches
on the two sides of the shell. Wherever ambiguous quantities are
encountered [such as in equations (\ref{dothatRdothatV}) below], the
equations are understood to hold individually on each
side of the shell.

The coordinates $(t,r)$ of section \ref{sec:metric} can be embedded in the
metric (\ref{EF-metric}) as $V = V(t,r)$ and $R=R(t,r)$,
independently on each side of the shell. We obtain
\begin{mathletters}
\label{g-of-VR}
\begin{eqnarray}
g_{tt} &=&
- F {\dot V}^2 - 2\eta {\dot V} {\dot R}
\ \ ,
\\
g_{rr} &=&
- F {V'}^2 - 2\eta {V'} {R'}
\ \ ,
\label{grr-of-VR}
\\
g_{tr} &=&
- F {\dot V} {V'} - \eta \left({\dot V} {R'} + {V'} {\dot R}\right)
\ \ .
\label{gtr-of-VR}
\end{eqnarray}
\end{mathletters}
As the surfaces of constant $t$ are spacelike, $\eta V' < 0$ everywhere.
Expressing the ADM variables in terms of the metric components and
using~(\ref{g-of-VR}), we find
\begin{equation}
{\eta N \over \Lambda} - N^r = -  { {\dot V} \over V'}
\ \ .
\label{VdotoverVprime}
\end{equation}

As in section~\ref{sec:metric}, the shell history is written as
$r=\zr(t)$, and we write $\hat R (t) := R\biglb(t,\zr(t)\bigrb)$. We also
write, independently on each side of the shell, $\hat V (t) :=
V\biglb(t,\zr(t)\bigrb)$, and similarly for $\widehat{\dot{R}}$, ${\widehat
{R'}}$, $\widehat{\dot{V}}$, ${\widehat {V'}}$, and so on. We then have,
independently on each side of the shell,
\begin{mathletters}
\label{dothatRdothatV}
\begin{eqnarray}
\dot{\hat R} &=& \widehat{\dot{R}} + \dot{\zr} {\widehat {R'}}
\ \ ,
\label{dothatR}
\\
\dot{\hat V} &=& \widehat{\dot{V}} + \dot{\zr} {\widehat {V'}}
\ \ .
\label{dothatV}
\end{eqnarray}
\end{mathletters}

With these preliminaries, we turn to the shell equations of motion. First,
equation (\ref{rhat-eom}) implies, with the help of (\ref{VdotoverVprime})
and~(\ref{dothatV}), that $\dot{\hat V}=0$. The shell history is therefore a
hypersurface of constant~$V$, independently on each side. These equations
also imply that $\widehat{\dot{V}} / {\widehat {V'}}$ is unambiguous and
\begin{equation}
\dot{\zr} = - \widehat{\dot{V}} / {\widehat {V'}}
\ \ .
\label{rrVdotoverVprime}
\end{equation}
Equations (\ref{grr-of-VR}) and (\ref{gtr-of-VR})
yield, after eliminating $\widehat{\dot{V}}$ and $\widehat{\dot{R}}$
with the help of (\ref{dothatR}) and~(\ref{rrVdotoverVprime}),
the relation
\begin{equation}
\eta {\widehat {V'}} \dot{\hat R} = - \dot{\zr} \widehat{g_{rr}} -
\widehat{g_{tr}}
\ \ .
\label{etaVR}
\end{equation}
As the right-hand side of (\ref{etaVR}) is unambiguous, and as $\dot{\hat
R}\ne0$, (\ref{etaVR}) implies that ${\widehat {V'}}$ is unambiguous. Thus,
both ${\widehat {V'}}$ and $\widehat{\dot{V}}$ are unambiguous.

Consider next the constraint~(\ref{delta-ham}). As ${\widehat {V'}}$
and ${\widehat {g_{rr}}}$ are unambiguous, equation
(\ref{grr-of-VR}) implies $\Delta R' = - \casehalf \eta {\widehat {V'}}
\Delta F$. Using~(\ref{EF-metric-F}), the constraint (\ref{delta-ham})
becomes
\begin{equation}
\zp = - {\widehat {V'}} \Delta M
\ \ .
\label{zp-of-DeltaM}
\end{equation}

Consider then equation~(\ref{rhatmom-eom}). Using (\ref{rrVdotoverVprime})
and the relation
${\dot {\widehat{V'}}} =
{\widehat{{\dot V}'}} +
\dot{\zr} \,
{\widehat{V''}}$,
a straightforward calculation yields
\begin{equation}
\left[ \left( { {\dot V} \over V'} \right)'\right]^{\outwidehat}
=
{ {\dot {\widehat{V'}}} \over {\widehat{V'}} }
\ \ .
\label{dothatprimeVoverhatprimeV}
\end{equation}
As the right-hand side of (\ref{dothatprimeVoverhatprimeV}) is unambiguous,
equations (\ref{VdotoverVprime}) and (\ref{dothatprimeVoverhatprimeV}) show
that the right-hand side of (\ref{rhatmom-eom}) is unambiguous. Further,
when (\ref{zp-of-DeltaM}) holds, it is seen that (\ref{rhatmom-eom}) is
identically satisfied.

What remains is equation~(\ref{delta-momR-eq-cov}). In the
coordinates (\ref{EF-metric}) we have, again
independently on the two sides of the shell,
${\widehat{v^a}} = (0,{\widehat{v^R}},0,0)$, and
\begin{equation}
\left[ v_b v^a \nabla_a (\partial_r)^b \right]^{\outwidehat}
=
- \eta \left[ {(v^R)}^2 \, (\partial_r)^V,\!{}_{R} \right]^{\outwidehat}
\ \ .
\label{delta-momR-cov2}
\end{equation}
As ${\widehat{v^R}}=\dot{\hat R}$, ${\widehat{v^R}}$ is
unambiguous. As ${\widehat{(\partial_r)^V}} = {\widehat{V'}} \,
{\widehat{(\partial_r)^r}}={\widehat{V'}}$, we see that
${\widehat{(\partial_r)^V}}$ is unambiguous. As the vector field denoted on
each side by $\partial/\partial R$ is continuous at the shell and tangent
to the shell history, $\left[(\partial_r)^V,\!{}_{R}\right]^{\outwidehat}$
is unambiguous. Therefore, the right-hand side of (\ref{delta-momR-cov2}) is
unambiguous, and equation (\ref{delta-momR-eq-cov}) is identically
satisfied.

\section{Equations of motion for a static shell}
\label{app:static-shell}

\begin{figure}
\begin{center}
\vglue 3 cm
\leavevmode
\epsfysize=6cm
\epsfbox{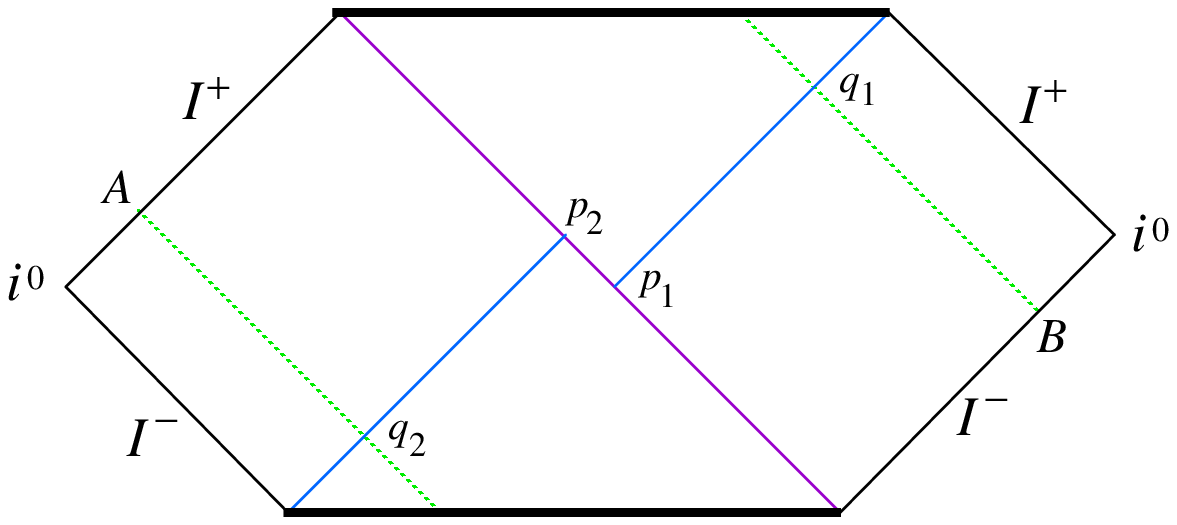}
\end{center}
\vskip 3 cm
\caption{The Penrose diagram for a spacetime in which the shell straddles a
horizon. The shell history is the line passing through points $p_1$
and~$p_2$. The spacetimes on the two sides share a common Schwarzschild
mass~$\bm$. The shell has been taken left-moving, which means $\eta=-1$.
The diagram corresponding to $\eta=1$ is obtained through time (or,
equivalently, space) inversion. The spacetime is uniquely determined by
the values of $\bm$ and~$\eta$. The points $A$ and $B$ and the dotted null
lines ending at them are as in figure~\ref{fig:nonhorizon}.}
\label{fig:horizon}
\end{figure}

In this appendix we verify the claims in the last paragraph of
appendix~\ref{app:dist-eoms-gen}. For concreteness, and without loss of
generality, we may take $\eta=-1$, so that the shell is moving to the left,
and the geometry is as in figure~\ref{fig:horizon}.

On each side of the shell, we introduce the Kruskal null coordinates,
\begin{mathletters}
\label{Kruskal-metric}
\begin{eqnarray}
ds^2 &=& - G  \, du \, dv
+ R^2 d\Omega^2
\ \ ,
\\
G &=& (32 M^3 / R) \exp(-R/2M)
\ \ ,
\label{Kruskal-metric-G}
\\
-uv &=& \left({R \over 2M} -1 \right) \exp(R/2M)
\ \ ,
\label{Kruskal-metric-uv}
\end{eqnarray}
\end{mathletters}
where $M$ is the common value of the Schwarzschild mass. When there is a
need to distinguish the two coordinate patches, we write the coordinates as
$(u_\pm,v_\pm)$, with the upper (lower) sign referring to the patch on the
right (left). The ranges of the coordinates are $v_+>0$ and $v_-<0$, with
$-\infty<u_\pm<\infty$, and the shell history lies at the common horizon at
$v_\pm=0$. When the index is suppressed, equations containing ambiguous
terms are understood to hold individually on each side of the shell.

Embedding the coordinates $(t,r)$ in the metric (\ref{Kruskal-metric}) as
$u = u(t,r)$ and $v=v(t,r)$, independently on each side of the shell, we
obtain
\begin{mathletters}
\label{g-of-uv}
\begin{eqnarray}
g_{tt} &=&
- G {\dot u} {\dot v}
\ \ ,
\\
g_{rr} &=&
- G {u'} {v'}
\ \ ,
\label{grr-of-uv}
\\
g_{tr} &=&
- \casehalf G \left({\dot u} {v'} + {u'} {\dot v} \right)
\ \ .
\label{gtr-of-uv}
\end{eqnarray}
\end{mathletters}
As the surfaces of constant $t$ are spacelike, $v' > 0$ everywhere.
Expressing the ADM variables in terms of the metric components and
using~(\ref{g-of-uv}), we find
\begin{equation}
{N \over \Lambda} + N^r = { {\dot v} \over v'}
\ \ .
\label{vdotovervprime}
\end{equation}

As above, we introduce the quantities $\hat u$, $\widehat{\dot{u}}$,
${\widehat {u'}}$, and so on, and similarly for~$v$. The counterparts of
equations (\ref{dothatRdothatV}) read
\begin{mathletters}
\begin{eqnarray}
\dot{\hat u} &=& \widehat{\dot{u}} + \dot{\zr} {\widehat {u'}}
\ \ ,
\label{dothatu}
\\
\dot{\hat v} &=& \widehat{\dot{v}} + \dot{\zr} {\widehat {v'}}
\ \ .
\label{dothatv}
\end{eqnarray}
\end{mathletters}
Equation (\ref{rhat-eom}) then implies, with the help
of (\ref{vdotovervprime}) and~(\ref{dothatv}), that
$\dot{\hat v}=0$, $\widehat{\dot{v}} / {\widehat {v'}}$ is unambiguous, and
\begin{equation}
\dot{\zr} = - \widehat{\dot{v}} / {\widehat {v'}}
\ \ .
\label{rrvdotovervprime}
\end{equation}

Consider next equation~(\ref{delta-momR-eq-cov}). In the
coordinates (\ref{Kruskal-metric}) we have,
independently on the two sides of the shell,
${\widehat{v^a}} = ({\widehat{v^u}},0,0,0)$, and
\begin{equation}
\left[ v_b v^a \nabla_a (\partial_r)^b \right]^{\outwidehat}
=
\left[ {(v^u)}^2 \, (\partial_r)_{u,u} \right]^{\outwidehat}
\ \ .
\end{equation}
Equation (\ref{delta-momR-eq-cov}) therefore reads
\begin{equation}
{(v^{u_+})}^2 \, (\partial_r)_{u_+,u_+}
=
{(v^{u_-})}^2 \, (\partial_r)_{u_-,u_-}
\ \ ,
\end{equation}
at the junction $v_\pm=0$. Writing
$v^{u_-}= (d{u_-}/d{u_+}) v^{u_+}$,
$(\partial_r)_{u_-}= (d{u_+}/d{u_-}) (\partial_r)_{u_+}$,
and
$\partial_{u_-} = (d{u_+}/d{u_-}) \partial_{u_+}$,
we obtain $d^2 {u_-} / d {u_+^2} = 0$.
This means that at the junction $v_\pm=0$ we have
\begin{equation}
{u_+}= \alpha u_- + \beta
\ \ ,
\label{affine-us}
\end{equation}
where $\alpha$ and $\beta$ are constants and $\alpha>0$. As the Kruskal
coordinates are affine parameters along the horizons, this means that the
soldering of the two geometries along the common horizon is affine. 

Consider next the constraint~(\ref{delta-ham}). {}From
(\ref{Kruskal-metric-uv}) we have
$\widehat{R'} = -2M \hat{u} \widehat{v'}$. The constraint (\ref{delta-ham})
therefore reads
\begin{equation}
\zp = -4 M^2 \Delta (u v')
\ \ .
\label{delta-ham-aa1}
\end{equation}
Equations (\ref{grr-of-uv}) and (\ref{gtr-of-uv}) yield, after eliminating
$\widehat{\dot{v}}$ and $\widehat{\dot{u}}$ with the help of
(\ref{dothatu}) and~(\ref{rrvdotovervprime}), the relation
\begin{equation}
8M^2 \, {\widehat {v'}} \, \dot{\hat u}
= - \dot{\zr} \widehat{g_{rr}} -
\widehat{g_{tr}}
\ \ .
\label{vprimedothatu}
\end{equation}
As the right-hand side of (\ref{vprimedothatu}) is unambiguous, ${\widehat
{v'}} \dot{\hat u}$ is unambiguous. The affine relation (\ref{affine-us})
implies $\dot{\widehat{u_+}} = \alpha \dot{\widehat{u_-}}$, and hence
\begin{equation}
\widehat{v_+'}
=
\alpha^{-1} \widehat{v_-'}
\label{hatvplusprimeis}
\ \ .
\end{equation}
Hence $\Delta
(u v') = \beta \widehat{v_+'}$, and equation (\ref{delta-ham-aa1})
takes the form
\begin{equation}
\zp = - 4 M^2 \beta  \, \widehat{v_+'}
\ \ .
\label{delta-ham-aa2}
\end{equation}
As $\zp<0$ by assumption and $\widehat{v_+'}>0$, we have $\beta>0$. This
means that the right-hand-side bifurcation two-sphere occurs earlier on the
history than the left-hand-side bifurcation two-sphere, as shown in
figure~\ref{fig:horizon}.

Consider finally equation~(\ref{rhatmom-eom}). Using
(\ref{rrvdotovervprime}) and proceeding as
with~(\ref{dothatprimeVoverhatprimeV}), we find
\begin{equation}
\left[ \left( { {\dot v} \over v'} \right)'\right]^{\outwidehat}
=
{ {\dot {\widehat{v'}}} \over {\widehat{v'}} }
\ \ .
\label{dothatprimevoverhatprimev}
\end{equation}
By~(\ref{hatvplusprimeis}), the right-hand side of
(\ref{dothatprimevoverhatprimev}) is unambiguous.
Equations (\ref{vdotovervprime}) and (\ref{dothatprimevoverhatprimev})
then show that the right-hand side of (\ref{rhatmom-eom}) is unambiguous.
When (\ref{delta-ham-aa2}) holds, it is seen that
(\ref{rhatmom-eom}) is identically satisfied.

\newpage

\newpage

\begin{references}

\bibitem{redmount-shell}
I.~H. Redmount,
Prog.\ Theor.\ Phys.\ {\bf 73}, 1401 (1985).

\bibitem{draythoft1}
T.~Dray and G. 't~Hooft,
Nucl.\ Phys.\ {\bf B253}, 173 (1985).

\bibitem{draythoft-cmp}
T.~Dray and G. 't~Hooft,
Commun.\ Math.\ Phys.\ {\bf 99}, 613 (1985).

\bibitem{clarkedray}
C.~J.~S. Clarke and T.~Dray,
Class.\ Quantum Grav.\ {\bf 4}, 265 (1987).

\bibitem{barr-is}
C.~Barrabes and W.~Israel,
Phys.\ Rev.\ D {\bf 43}, 1129 (1991).

\bibitem{bcmn}
B.~K. Berger,
D.~M. Chitre,
V.~E. Moncrief, and
Y.~Nutku,
Phys.\ Rev.\ D {\bf 8}, 3247 (1973).

\bibitem{unruh}
W.~G. Unruh,
Phys.\ Rev.\ D {\bf 14}, 870 (1976).

\bibitem{thiemann1}
T.~Thiemann and H.~A. Kastrup,
Nucl.\ Phys.\ {\bf B399}, 211 (1993).
(gr-qc/9310012)

\bibitem{thiemann2}
H.~A. Kastrup and T.~Thiemann,
Nucl.\ Phys.\ {\bf B425}, 665 (1994).
(gr-qc/9401032)

\bibitem{kuchar1}
K.~V. Kucha\v{r},
Phys.\ Rev.\ D {\bf 50}, 3961 (1994).
(gr-qc/9403003)

\bibitem{thiemann3}
T.~Thiemann,
Int.\ J.\ Mod.\ Phys.\ D {\bf 3}, 293 (1994).

\bibitem{thiemann4}
T.~Thiemann,
Nucl.\ Phys.\ {\bf B436}, 681 (1995).

\bibitem{LW2}
J.~Louko and
B.~F. Whiting,
Phys.\ Rev.\ D {\bf 51}, 5583 (1995).
(gr-qc/9411017)

\bibitem{cava1}
M.~Cavagli\`a, V. de~Alfaro, and A.~T. Filippov,
Int.\ J.\ Mod.\ Phys.\ D {\bf 4}, 661 (1995).
(gr-qc/9411070)

\bibitem{cava2}
M.~Cavagli\`a, V. de~Alfaro, and A.~T. Filippov,
Int.\ J.\ Mod.\ Phys.\ D {\bf 5}, 227 (1996).
(gr-qc/9508062)

\bibitem{lau1}
S.~R. Lau,
Class.\ Quantum Grav.\ {\bf 13}, 1541 (1996).
(gr-qc/9508028)

\bibitem{marolf-boundary}
D.~Marolf,
Class.\ Quantum Grav.\ {\bf 13}, 1871 (1996).
(gr-qc/9510023)

\bibitem{lm}
J.~Louko and J.~M\"akel\"a,
Phys.\ Rev.\ D {\bf 54}, 4982 (1996).
(gr-qc/9605058)

\bibitem{lou-win}
J.~Louko and S.~N. Winters-Hilt,
Phys.\ Rev.\ D {\bf 54}, 2647 (1996).
(gr-qc/9602003)

\bibitem{kuns1}
J.~Gegenberg and G.~Kunstatter,
Phys.\ Rev.\ D {\bf 47}, R4192 (1993).
(gr-qc/9302006)

\bibitem{kuns2}
J.~Gegenberg, G.~Kunstatter,
and D.~Louis-Martinez,
Phys.\ Rev.\ D {\bf 51}, 1781 (1995).
(gr-qc/9408015)

\bibitem{kuns3}
D.~Louis-Martinez and G.~Kunstatter,
Phys.\ Rev.\ D {\bf 52}, 3494 (1995).
(gr-qc/9503016)

\bibitem{bro-kiefer}
T.~Brotz and C.~Kiefer,
Phys.\ Rev.\ D {\bf 55}, 2186 (1997).
(gr-qc/9608031)

\bibitem{euclidean-refs}
R.~Laflamme,
in {\it Origin and Early History of the Universe:
Proceedings of the 26th Li\`ege International
Astrophysical Colloquium (1986)\/},
edited by J.~Demaret
(Universit\'e de Li\`ege,
Institut d'Astro\-physique, 1987);
R.~Laflamme, 
Ph.D. thesis (University of Cambridge,1988);
B.~F. Whiting and J.~W. York,
Phys.\ Rev.\ Lett.\ {\bf 61}, 1336 (1988);
H.~W. Braden,
J.~D. Brown,
B.~F. Whiting, and
J.~W. York,
Phys.\ Rev.\ D {\bf 42}, 3376 (1990);
J.~J. Halliwell and
J.~Louko,
Phys.\ Rev.\ D {\bf 42}, 3997 (1990);
G.~Hayward and
J.~Louko,
Phys.\ Rev.\ D {\bf 42}, 4032 (1990);
J.~Louko and
B.~F. Whiting,
Class.\ Quantum Grav.\ {\bf 9}, 457 (1992);
J.~Melmed and B.~F. Whiting,
Phys.\ Rev.\ \rm D {\bf 49}, 907 (1994);
S.~Carlip and C.~Teitelboim,
Class.\ Quantum Grav.\ {\bf 12}, 1699 (1995)
(gr-qc/9312002);
S.~Carlip and C.~Teitelboim,
Phys.\ Rev.\ D {\bf 51}, 622 (1995)
(gr-qc/9405070);
G.~Oliveira-Neto,
Phys.\ Rev.\ D {\bf 53}, 1977 (1996).

\bibitem{frolov}
V.~P. Frolov,
Zh.\ Eksp.\ Teor.\ Fiz.\ {\bf 66}, 813 (1974)
[Sov.\ Phys.\ JETP {\bf 39}, 393 (1974)].

\bibitem{berezin88}
V.~A. Berezin,
N.~G. Kozmirov,
V.~A. Kuzmin,
and
I.~I. Tkachev,
Phys.\ Lett.\ B {\bf 212}, 415 (1988).

\bibitem{fischler}
W.~Fischler,
D.~Morgan, and
J.~Polchinski,
Phys.\ Rev.\ D {\bf 42}, 4042 (1990).

\bibitem{hajicek}
P.~H\'aj\'{\i}\v{c}ek,
Commun.\ Math.\ Phys.\ {\bf 150}, 545 (1992).

\bibitem{HKK}
P.~H\'aj\'{\i}\v{c}ek,
B.~S. Kay,
and
K.~V. Kucha\v{r},
Phys.\ Rev.\ D {\bf 46}, 5439 (1992).

\bibitem{kraus-wilczek1}
P.~Kraus and F.~Wilczek,
Nucl.\ Phys.\ {\bf B433}, 403 (1995).
(gr-qc/9408003)

\bibitem{kraus-wilczek2}
P.~Kraus and F.~Wilczek,
Nucl.\ Phys.\ {\bf B437}, 231 (1995).
(hep-th/9411219)

\bibitem{nakamura}
K.~Nakamura,
Y.~Oshiro, and
A.~Tomimatsu,
Phys.\ Rev.\ D {\bf 53}, 4356 (1996).
(gr-qc/9506032)

\bibitem{dolgov}
A.~D. Dolgov and
I.~B. Khriplovich,
Phys.\ Lett.\ B {\bf 400}, 12 (1997). 
(hep-th/9703042)

\bibitem{vakkuri-kraus}
E.~Keski-Vakkuri and P.~Kraus,
Nucl.\ Phys.\ {\bf B491}, 249 (1997).
(hep-th/9610045)

\bibitem{bicak-haji}
P.~H\'aj\'{\i}\v{c}ek 
and
J.~Bi\v{c}\'{a}k, 
Phys.\ Rev.\ D {\bf 56}, 4706 (1997).
(gr-qc/9706022)

\bibitem{kol-eard}
S.~J. Kolitch and D.~M. Eardley,
Phys.\ Rev.\ D {\bf 56}, 4663 (1997).
(gr-qc/9706033)

\bibitem{FriLoWi}
J.~L. Friedman,
J.~Louko, and
S.~N. Winters-Hilt,
Phys.\ Rev.\ D {\bf 56}, 7674 (1997).
(gr-qc/9706051)

\bibitem{ansoldi+}
S.~Ansoldi, A.~Aurilia, R.~Balbinot, and 
E.~Spallucci, 
Class.\ Quantum Grav.\ {\bf 14}, 2727 (1997). 
(gr-qc/9706081)

\bibitem{lund}
F.~Lund,
Phys.\ Rev.\ D {\bf 8}, 3247 (1973).

\bibitem{romano1}
J.~D. Romano,
``Spherically Symmetric Scalar Field Collapse: An Example of the Spacetime
Problem of Time," Report UU-REL-95/1/13, gr-qc/9501015.

\bibitem{romano2}
J.~D. Romano,
Phys.\ Rev.\ D {\bf 55}, 1112 (1997).

\bibitem{GT}
R.~Geroch and J.~Traschen,
Phys.\ Rev.\ D {\bf 36}, 1017 (1987).

\bibitem{painleve}
P.~Painlev\'e,
C.~R. Acad.\ Sci.\ (Paris) {\bf 173}, 677 (1921).

\bibitem{gullstrand}
A.~Gullstrand,
Ark.\ Mat.\ Astron.\ Fys.\ {\bf 16}(8), 1 (1922).

\bibitem{israel-flat}
W.~Israel,
in {\it Three Hundred Years of Gravitation,}
edited by S.~W. Hawking and W.~Israel
(Cambridge University Press, Cambridge, 1987),
p.~234.

\bibitem{israel-shell}
W.~Israel,
Nuovo Cimento {\bf B44}, 1 (1966).

\bibitem{MTW}
C.~W. Misner, K.~S. Thorne, and J.~A. Wheeler,
{\it Gravitation\/} (Freeman, San Francisco, 1973).

\bibitem{teitel-dirac}
C.~Teitelboim,
Ann.\ Phys.\ (N.Y.) {\bf 79}, 524 (1973).

\bibitem{henn-teit-book}
M.~Henneaux and C.~Teitelboim,
{\it Quantization of Gauge Systems\/}
(Princeton University Press, Princeton, NJ, 1992).

\bibitem{beig-om}
R.~Beig and N.~\'O~Murchada,
Ann.\ Phys.\ (N.Y) {\bf 174}, 463 (1987).

\bibitem{topocen}
J.~L. Friedman, K.~Schleich, and D.~M. Witt,
Phys.\ Rev.\ Lett.\ {\bf 71}, 1486 (1993);
Erratum,
Phys.\ Rev.\ Lett.\ {\bf 75}, 1872 (1995).
(gr-qc/9305017)

\bibitem{haji-kijo1}
P.~H\'aj\'{\i}\v{c}ek and J.~Kijowski,
Phys.\ Rev.\ D {\bf 57}, 914 (1998).
(gr-qc/9707020) 

\bibitem{haji-gensshell}
P.~H\'aj\'{\i}\v{c}ek, 
Phys.\ Rev.\ D {\bf 57}, 936 (1998).
(gr-qc/9708008) 

\bibitem{haji-kijo2}
P.~H\'aj\'{\i}\v{c}ek and J.~Kijowski,
private communication.

\bibitem{KRV}
K.~V. Kucha\v{r}, J.~D. Romano, and M.~Varadarajan,
Phys.\ Rev.\ D {\bf 55}, 795 (1997).
(gr-qc/9608011)


\end{references}
\end{document}